\RequirePackage{fix-cm}
\documentclass[twocolumn,epjc3]{svjour3}  
\smartqed  
\RequirePackage{graphicx}
\RequirePackage{amsmath}
\RequirePackage{amsfonts}
\RequirePackage{mathrsfs}
\usepackage{hyperref,booktabs}
\usepackage{subcaption}
\usepackage{xcolor,amsfonts}
\usepackage{float}
\usepackage{lineno}
\journalname{Eur. Phys. J. A}

\emergencystretch=\maxdimen
\begin{document}
\title{Polarized Target Nuclear Magnetic Resonance Measurements with Deep Neural Networks
}


\author{D. Seay
        \and
        I. P. Fernando 
        \and 
        D. Keller
}

\thankstext{t1}{This work was supported by the Department of Energy (DOE), United States of America contract DE-FG02-96ER40950.}


\institute{Department of Physics, University of Virginia, Charlottesville, Virginia 22904, USA}

\date{Received: date / Accepted: date}

\maketitle

\begin{abstract}
Continuous-wave Nuclear Magnetic Resonance (CW-NMR) operated in constant-current mode has served as a foundational technique for polarization measurement in solid-state dynamically polarized targets within nuclear and high-energy physics experiments for several decades, and it remains an essential tool. Conventional Q-meter-based phase-sensitive detection is critical for precise real-time determination of target polarization during scattering runs. However, the accuracy and reliability of these measurements are frequently compromised by elevated noise levels, baseline drift, and systematic uncertainties arising from signal isolation and fitting, ultimately degrading the overall experimental figure of merit.
In this work, we report the first successful application of neural network architectures to continuous-wave NMR polarization metrology. By leveraging advanced machine learning techniques for signal extraction and denoising, we achieve a substantial reduction of fitting uncertainties under a variety of realistic simulated and experimental conditions. These improvements translate directly into more robust real-time (online) polarization monitoring and higher precision in subsequent offline analysis. By reducing analysis-induced uncertainty, the resulting methodology can improve the effective figure of merit for scattering experiments employing dynamically polarized targets and provides a new toolset for NMR-based polarimetry in high-energy and nuclear physics.
\end{abstract}

\section{Introduction}
\label{intro}

Continuous-wave Nuclear Magnetic Resonance (CW-NMR) remains the method of choice for real-time polarization monitoring in solid-state dynamically polarized targets employed in nuclear and high-energy physics scattering experiments. The standard implementation relies on resonant Q-meters operated in constant-current mode, which facilitate phase-sensitive detection of the absorptive and dispersive components of the RF susceptibility\cite{Court1993}. Although this approach is mature and widely fielded, its performance under realistic experimental conditions is often fundamentally constrained by multiple degradation mechanisms: baseline distortions, abrupt detuning caused by cable discontinuities or mechanical perturbations of the resonant circuits, radio-frequency (RF) interference inherent to experimental environments, and limitations on clean power. Consequently, conventional polarization extraction techniques---whether based on thermal-equilibrium (TE) calibration scaling or analytic lineshape fitting---can incur relative uncertainties of several percent arising solely from fitting systematics. At low polarization levels, where the NMR signal amplitude becomes comparable to the prevailing noise floor, stochastic fluctuations and measurement-to-measurement variability further degrade both accuracy and precision. Even under optimal conditions, the widely used Liverpool Q-meter (described in Sec.~\ref{sec:qmeter}) exhibits an intrinsic relative uncertainty of approximately 1\% \cite{Court1993}.
The increasing demand for accurate polarization measurements across a broad spectrum of facilities, from high-luminosity hadron and lepton beamlines to various meson and photon beam experiments, underscores the need for measurement methods that remain robust against baseline skew, tune drift, elevated noise, and deliberate RF field modulation. Modern machine learning techniques, particularly deep artificial neural networks (ANNs), are ideally suited to address some of these challenges. By training on large, physically accurate ensembles of simulated CW-NMR signals that systematically incorporate variations in circuit parameters, environmental noise, baseline morphology, and target-specific lineshapes, neural networks can learn highly nonlinear mappings from raw complex voltage and phase to polarization with full uncertainty quantification. This approach eliminates the need for real-time analytic fitting during online monitoring, enhances resilience to RF artifacts, and preserves or improves ultimate accuracy and precision for both spin-1/2 and spin-1 systems.

In this work, we report the first comprehensive application of deep neural networks to polarization metrology using Q-meter-based CW-NMR. For spin-1/2 targets, we introduce an area-integration methodology that yields stable and unbiased polarization estimates even in the presence of severe baseline shifts and noise contamination. For spin-1 systems exhibiting a Pake doublet (characteristic of deuterated targets in non-cubic lattices), we benchmark direct vector-polarization extraction and develop the lineshape-based framework needed for future tensor-polarization extraction. We present detailed performance benchmarks, compare the robustness and uncertainty budgets of the respective methods, validate against conventional analyses on experimental data, and discuss pathways toward real-time deployment and further refinement.

The remainder of this paper is organized as follows. Section~\ref{sec:qmeter} summarizes the Liverpool Q-meter system, the RF circuit model, and the dominant noise and baseline effects that motivate the simulation framework. Section~\ref{sec:theory} reviews the spin--1 NMR lineshape relations needed to generate training data for deuterated targets. Section~\ref{sec:historical} summarizes relevant historical extraction methods and their fitting-related uncertainties. Section~\ref{sec:methodology} introduces the conventional and DNN-based extraction strategies and defines the analysis metrics. Section~\ref{sec:architecture} describes the model architectures and training procedure. Section~\ref{sec:results} presents the benchmark results and comparisons with conventional TE and lineshape fitting methods. Section~\ref{sec:tensor} discusses how the same tools can be extended to tensor-enhanced measurements, and Section~\ref{sec:conclusion} summarizes the paper and outlines future directions.
\section{Liverpool Q-meter System}
\label{sec:qmeter}

Polarization measurements of the target sample are performed using a system equipped with a Liverpool Q-meter \cite{Court1993}. The Q-meter operates in conjunction with a data-acquisition system to record the voltage at each frequency step within its operational range of $3$--$300$~MHz, enabling measurements of materials whose Larmor frequencies fall within this bandwidth.

The target signal is carried through a special copper coaxial transmission line whose total electrical length is constructed to be an integer multiple of $\lambda/2$, where $\lambda$ is the wavelength corresponding to the Larmor frequency of the target material in the cable (i.e., scaled by the cable's velocity factor). Because a transmission line reproduces its input impedance every $\lambda/2$, using an integer number of half-wavelength sections ensures that the phase of the detected NMR signal is preserved and that the resonant lineshape at the Q-meter remains undistorted.  It is important to note that the 
$n \lambda/2$ cable condition is exact only at the center frequency. As the frequency is swept
away from resonance, reactive components of the circuit and cable alter the imaged impedance,
producing a larger Q-curve background. For typical coaxial cables with a velocity factor of approximately $0.78$ (for cables with a foam/PTFE dielectric), the electrical half-wavelength is approximately $55\ \text{cm}$ at $213\ \text{MHz}$ (the proton Larmor frequency at $5~\text{T}$) and approximately $360\ \text{cm}$ at $32.7\ \text{MHz}$ (the central deuteron Larmor frequency at $5~\text{T}$). Accurate knowledge of this electrical length is essential: even modest deviations from an exact $n\,\lambda/2$ configuration introduce shifts in the circuit's Q-curve that appear as left- or right-leaning asymmetries in the continuous-wave NMR baseline.

In the Liverpool Q-meter~\cite{Court1993} and its descendants, the current through the target coil is held approximately constant, while the complex voltage developed across the low-impedance tuned circuit is measured. An RF phase-sensitive detector, referenced to the coil current itself, extracts the \emph{real part} of this voltage ($V_R$). This real-part signal is directly proportional to the total effective series resistance of the resonant circuit. The absorptive component of the nuclear susceptibility $\chi''(\omega)$ (whether from energy-absorbing or energy-emitting spin transitions) therefore manifests as a small positive or negative change in circuit loss, appearing as a proportional variation in $V_R$ that is essentially linear with polarization over a wide dynamic range. By contrast, the dispersive component $\chi'(\omega)$ primarily causes a slight detuning of the resonant frequency; because the measurement is performed with a fixed-frequency sweep rather than by tracking the resonance, $\chi'$ does not produce an independent orthogonal dispersion channel but instead contributes to the characteristic sloping Q-curve background. The Q-meter design configuration prioritizes the real part of the output voltage---rather than magnitude and phase or separate absorption and dispersion quadratures---which significantly improves linearity (nonlinearity typically below 0.2--1\%) and strongly suppresses the amplitude of the Q-curve background. The primary polarization signal is carried almost exclusively by the real-part (absorption-dominated) channel, with residual dispersive effects treated as baseline contributions to be removed by careful tuning, cable-length optimization, temperature stabilization, or modern subtraction and modeling techniques. Tuning ensures that the Q-meter operates optimally for a given experimental configuration. The $\lambda/2$ transmission line must be set to the correct electrical length to ensure proper system performance, while a variable capacitor is adjusted to balance the parallel inductor-capacitor (LC) tank of the circuit and achieve impedance matching between the Q-meter reference circuit and the inductive coil containing the target sample. In addition, phase tuning is performed to minimize reactive components, thereby isolating the real (in-phase) component of the signal. The imaginary (quadrature) component is separated by a $\pi/2$ phase shift, improving the fidelity of phase-sensitive detection. We note here that being out of tune with either the cable length or the phase frequently leads to errors in the extraction, especially when these changes are due to spontaneous jumps from mechanical shifts during data acquisition. These shifts can be corrected using a lineshape model that includes the off-tune response, but without correction they can contribute more than an additional 1\% relative error.

Standard polarization measurement techniques employing Q-meter-based NMR systems in scattering experiments typically achieve relative uncertainties at the level of 3--5\% \cite{Keller2020}. These uncertainties are dominated by limitations in polarization calibration, most notably those associated with thermal-equilibrium measurements (TEs), and by fitting-related uncertainties. Additional contributions arise from magnetic-field inhomogeneity, temperature drifts in the readout electronics, imperfect knowledge of the coil filling factor, 3D distribution of polarization in the sample, baseline distortions, statistical fluctuations, and limitations in pressure or temperature measurement and calibration. The dominance of these individual error sources is discussed in Sec.~\ref{sec:errors}.

Neural networks provide a powerful framework for improving both the accuracy and precision of polarization measurements extracted from continuous-wave NMR signals, primarily through the reduction of fitting-related uncertainties and, to a more limited extent, certain systematic components. By learning highly nonlinear mappings directly from raw detector outputs, these models can be trained to minimize the full polarization covariance across a wide range of experimental conditions. Moreover, their ability to generalize beyond the training data enables partial mitigation of some instrumental limitations (jumps in baseline and tune shifts) that are difficult to address using traditional analysis techniques.

In the specific context of the Liverpool Q-meter, the observed baseline shape, signal lineshape, and amplitude-to-polarization relationship are all influenced by subtle, interdependent instrumental effects. Traditional analytical calibration is capable of reaching the lower 1\% uncertainty bound but only under optimal conditions. A suitably trained neural network, by contrast, can implicitly capture these complex interactions and correct for them, provided it is trained on realistic data that encodes the full range of instrumental characteristics. To achieve this, we require high-quality training data generated from simulations that capture all relevant dependencies and the leading contributions to the full covariance.  In this regard, the simulated training data are highly tuned and matched to the experimental data with the added benefit that the true lineshape parameters and polarizations are known exactly.

To generate such training data, an accurate circuit model of the entire Q-meter system---including the oscillator dynamics, transmission-line behavior, probe tuning characteristics, and known sources of systematic error---was implemented (see Sec.~\ref{sec:errors} for a complete inventory of the modeled effects). By sampling this model over the relevant parameter space, we produced large, physically realistic datasets of the Q-meter circuit behavior (baseline or Q-curve).  We then add the signal lineshape, various noise components, and tuned or off-tune circuit variations, and perform Monte Carlo (MC) sampling to produce experimentally realistic baseline and polarization data for high-statistics training and testing.  With highly tuned MC data, the models can be trained to handle moderate residual baseline distortions, phase-mixing effects, and Q-meter-driven false asymmetries.

\subsection{Circuit Theory}
\label{sec-circuit}
We develop a Python-based Q-meter circuit simulation based on the Mathcad program originating from the Liverpool group \cite{CourtNMRSimulation}.
The Q-meter--based NMR system couples the target material to a sensing coil of intrinsic inductance $L_0$. Whether the coil is wound around the sample or embedded within it, its bare inductance remains fixed. Introducing the sample alters the coil's effective inductance through its magnetic susceptibility and coupling to the RF field, thereby shifting the circuit's resonance. We can express this modified inductance of the combined sample--coil system as
\begin{equation}
    L(\omega) = L_0 (1 + 4\pi\eta\chi(\omega))
\end{equation}
where $\omega$ is the probe frequency, $\eta$ is the filling constant of the coil, and $\chi(\omega)$ is the complex RF susceptibility, which can be expressed as $\chi(\omega) = \chi'(\omega) - i\chi^{\prime\prime}(\omega)$. This complex RF susceptibility has nonzero values at frequencies near the resonant frequency of a target nucleon. 
\begin{figure}
    \centering
    \includegraphics[width = 1.0\linewidth]{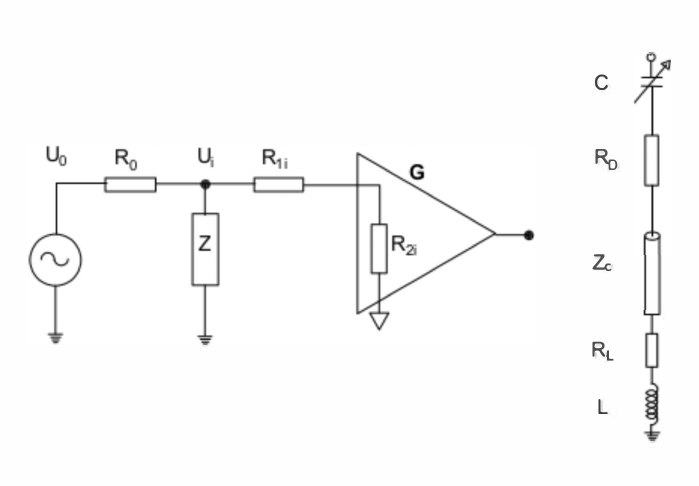}
    \caption{Schematic of the Q-meter circuitry\cite{Pentilla1998}.}
    \label{fig:Q-meter}
\end{figure}
The real part of the voltage across the Q-meter can be expressed as
\begin{equation}
    \operatorname{Re}\{u_i\} = \frac{U}{R_0} \left(\frac{\operatorname{Re}\{Z\} + Y[\operatorname{Re}\{Z\}^2 + \operatorname{Im}\{Z\}^2]}{[(1 + Y \operatorname{Re}\{Z\})^2 + Y^2 \operatorname{Im}\{Z\}^2]}\right)
\end{equation}
where $Y = \frac{1}{R_i} + \frac{1}{R_0}$ is the coupling admittance of the resonator, and $R_i$ is the total amplifier impedance, expressed as $R_i = R_{1i} + R_{2i}$ with $R_{1i}$ assumed to be purely resistive. Here, $R_0$ is the current-limiting resistance. A circuit diagram of the Q-meter is shown in Fig.~\ref{fig:Q-meter}, where $G$ is the amplifier gain, $R_D$ is the damping resistance, $C$ is the tuning capacitor, and $R_L$ is the inductive resistance. To avoid ambiguity, we denote the tuning-capacitor impedance by $Z_{\mathrm{cap}}(\omega)$ and the characteristic impedance of the transmission line by $Z_{\ell}$. In experiment, the Q-meter is tuned to detect the real part of the target signal, so the impedance of the resonant part of the circuit, $Z_T$, is written as
\begin{equation}
    Z_T = R_D + Z_{\mathrm{cap}}(\omega) + Z_{\ell} \left( \frac{Z_L + Z_{\ell}\tanh{\gamma l}}{Z_{\ell} + Z_L\tanh{\gamma l}}\right),
\end{equation}
where $C$ is the tuning capacitance (parameterized below as $C_{\mathrm{knob}}$), $R_D$ is the damping resistor, $Z_L$ is the coil impedance, and $l = n\lambda/2$ is the cable length for integer $n$. We can express the coil impedance as
\begin{equation}
\label{eq:impedance}
\begin{aligned}
Z_L^{(\chi)}(\omega)
&= R_L + i\omega L \left[ 1 + \eta_L\!\left(\chi^{\prime}(\omega) - i\chi^{\prime\prime}(\omega)\right) \right], \\[6pt]
Z_L^{(0)}(\omega)
&= \frac{\bigl(R + i\omega L\bigr)\left(\dfrac{1}{i\omega C_{\mathrm{stray}}}\right)}
{\bigl(R + i\omega L\bigr) + \left(\dfrac{1}{i\omega C_{\mathrm{stray}}}\right)} .
\end{aligned}
\end{equation}
Here $Z_L^{(\chi)}$ denotes the susceptibility-modified inductive branch used when the NMR response is included, while $Z_L^{(0)}$ denotes the circuit-only terminal impedance including the stray capacitance. The parameter $\eta_L$ is the filling factor, and $\chi^{\prime}(\omega)-i\chi^{\prime\prime}(\omega)$ is the complex magnetic susceptibility of the target material. In the second line, $R$ denotes the effective series resistance associated with the inductive branch of the resonant circuit, incorporating both the coil resistance and additional dissipative losses present at the coil terminals. The parameter $C_{\mathrm{stray}}$ represents the parasitic capacitance arising from the surrounding circuit environment. In Eq.~\ref{eq:impedance}, $Z_L$ is understood to be the appropriate coil impedance for the calculation being performed.
The impedance of the tuning capacitor is given by
\begin{equation}
Z_{\mathrm{cap}}(\omega) = \frac{1}{i\omega C(\omega)} .
\end{equation}
We can then express the total impedance of the Q-meter system, $Z(\omega)$, as
\begin{equation}
Z(\omega) =
\frac{R_1}{1 + \dfrac{R_1}{r + Z_{\mathrm{cap}}(\omega) + Z_T(\omega)}},
\end{equation}
where $R_1$ denotes the current-limiting (source) resistance of the RF drive stage, and $r$ represents the remaining series resistance of the resonant circuit. The propagation constant of the $\lambda/2$ transmission line can then be expressed as
\begin{align}
    \qquad \qquad \gamma &= \sqrt{(R_c + i\omega L_c)(G_c + i\omega C_c)} \nonumber \\ & \cong i\omega\sqrt{L_c C_c}\left(1 + \frac{1}{2iQ_c}\right).
\end{align}

The characteristic impedance of the coaxial line is defined as
\begin{equation}
    \qquad Z_{\ell} = \sqrt{\frac{R_c + i\omega L_c}{G_c + i\omega C_c}} \cong Z_0 \left(1 + \frac{1}{2iQ_c}\right),
\end{equation}
where the subscript $c$ denotes cable parameters, $Z_0 = \sqrt{L_c / C_c}$, and $Q_c = \omega L_c / R_c$. 

The Q-meter baseline curve was simulated by implementing this mathematical description of the Q-meter circuit 
\cite{CourtNMRSimulation} in various test configurations matched to experimental tests using a real Q-meter system. This implementation allows us to vary eight simulated parameters that characterize the Q-meter RF circuitry and environment. These parameters are:

\begin{itemize}
    \item $C_{\mathrm{knob}}$: the tuning capacitance in the circuit. In practice, this parameter is adjusted by turning a variable capacitor knob on the Q-meter card. $C_{\mathrm{knob}}$ is adjusted such that the baseline curve is tuned. It has units of pF and typically ranges from $0.01$--$100$~pF.
    \item $U$: input voltage. $U$ is used to calculate the operating current through the Q-meter circuit via the equation $I = U / R$, the impedance $Z(\omega)$, and the phase $\Phi_{\mathrm{total}}(\omega)$, which leads to $V(\omega) = IZ(\omega)e^{i\Phi_{\mathrm{total}}(\omega)}$. It is in units of V and typically ranges from $0.1$--$1$~V.
    \item $n/2$: the length of the $\lambda/2$ cable, also called the trim. For optimal performance, the cable length is configured in discrete multiples of $\lambda/2$ in order for the circuit to be at resonant frequency $\omega_0$. This makes the full cable length $n\lambda/2$. 
    \item $\eta_{L}$: filling factor of the coil. This factor is the level of coupling of spins in the target material to the sampling coil. It is unitless and ranges from $0$--$1$. 
    \item $C_{\mathrm{stray}}$: stray capacitance in the circuit environment, including parasitic capacitance between nearby electronic components or conductors. We can calculate the true capacitance of the LC circuit that is controlled by the variable capacitor knob to be
    \begin{equation}
        C(\omega) = k\times C_{\mathrm{knob}}
    \end{equation}
    The effective capacitance (including stray) can have a wide range of values, from a few to a few hundred pF, depending on the Larmor frequency.
    \item $\phi_0$: Phase offset. We model the slowly varying phase response in radians as
    \begin{equation}
        \Phi_{\mathrm{base}}(\omega) = a\omega^2 + b\omega + \phi_0,
    \end{equation}
    where $a$ and $b$ are constants. The offset $\phi_0$ is typically sampled within $[0,2\pi]$, but it can extend beyond that range modulo $2\pi$. A quadratic approximation is used to model the phase behavior within the waveguide. Higher-order terms are neglected because their contributions are small compared to other dominant sources of uncertainty and would be obscured by overlapping errors.
    \item $L_{0}$: coil inductance. This is typically around the value of $50$ nH and depends on the size of the NMR coil.
    \item DC offset: This is used in the simulations to account for any physical DC offset that cannot be immediately identified from experimental measurements. Although it is usually small, somewhere on the scale of $\approx 10^{-2}$ V, this can depend on the cable length.
\end{itemize}

Additionally, the frequency-dependent trim contribution to the phase is parameterized as
\begin{align}
    \text{slope}_{\phi} &= \frac{\delta\phi}{ \text{(Sweep Length)} \times 2\pi \times 10^6}, 
\end{align}
where $\delta\phi$ is a circuit-dependent phase constant. For example, for an NH$_3$ target at $5~\mathrm{T}$, whose proton Larmor frequency is near $213~\mathrm{MHz}$, a sweep length of $4~\mathrm{MHz}$ would cover approximately $209$--$217~\mathrm{MHz}$. The trim contribution is then
\begin{equation}
    \Phi_{\mathrm{trim}}(\omega) = \text{slope}_{\phi} \times (\omega - \omega_{0}),
\end{equation}
where $\omega_0$ is the resonant frequency being probed. The total phase, in radians, is
\begin{equation}
    \Phi_{\mathrm{total}}(\omega) = \Phi_{\mathrm{base}}(\omega) + \Phi_{\mathrm{trim}}(\omega).
\end{equation}
We finally describe the output voltage of the Q-meter as
\begin{equation}
    V_{\mathrm{out}}(\omega) = \Re\!\left[I Z_{\mathrm{total}}(\omega)e^{i\Phi_{\mathrm{total}}(\omega)}\right]
\end{equation}
where $I$ denotes the RF drive current applied to the resonant circuit. Maintaining a constant drive current is essential, as the spin-transition rate and steady-state response depend on the RF field amplitude, which is directly proportional to 
$I$. Temporal variations in the applied current would lead to time-dependent excitation conditions, inducing nonuniform spin transitions and perturbing the spin population away from the desired steady state. The Q-meter-based NMR design therefore balances high-fidelity polarization measurement with minimal perturbation of the enhanced spin polarization, ensuring that the measurement remains effectively non-destructive.

\begin{figure}[!h]
        \centering
        \includegraphics[width=1\linewidth]{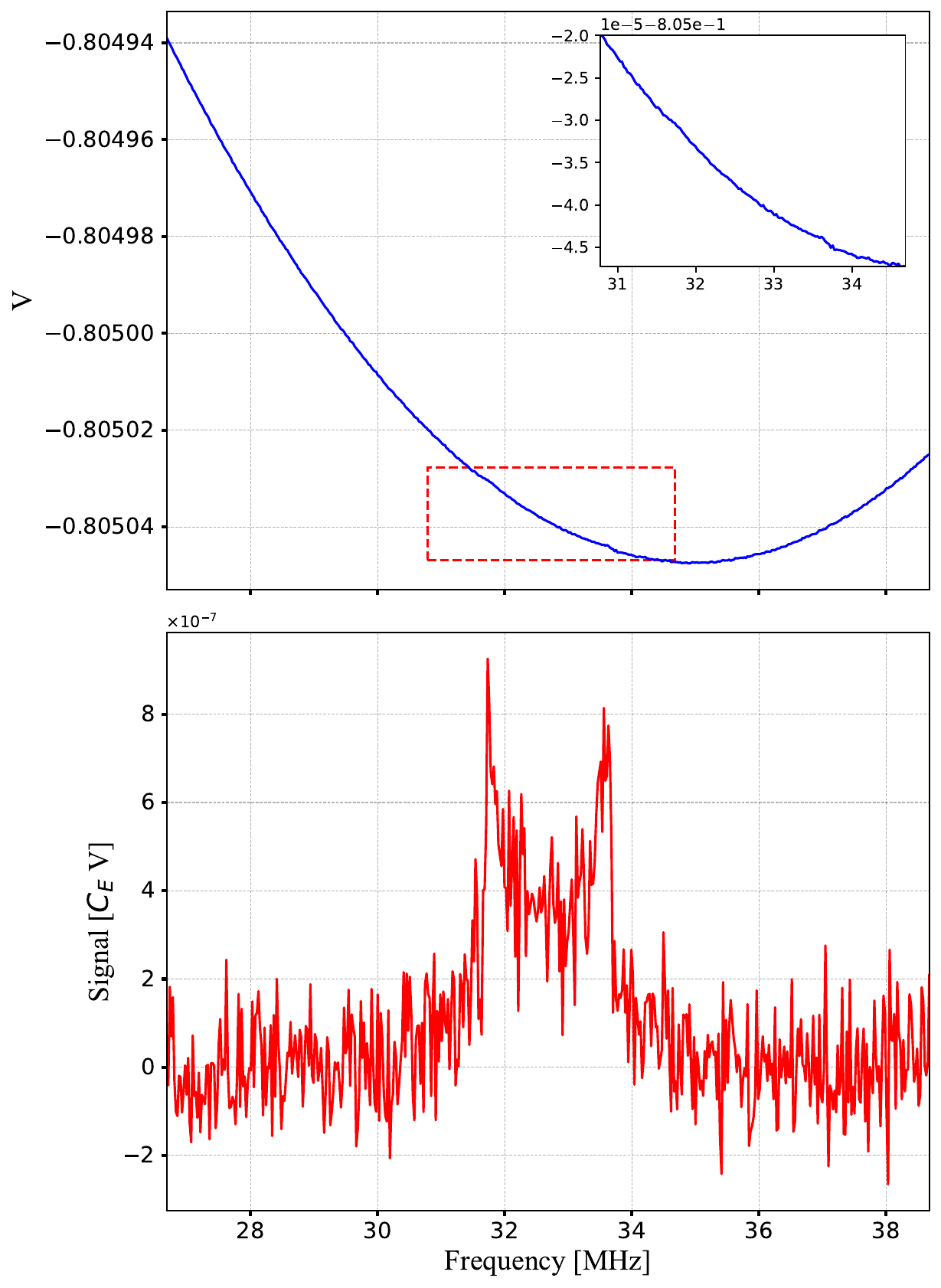}
        \caption{Top: Example of a simulated Q-meter baseline containing a deuteron NMR lineshape at $0.05\%$ polarization (TE), with MC noise added at a signal-to-noise ratio of 2.6. For clarity, the signal shown in the inset is scaled by a factor of $10^{2}$.
Bottom: The same signal as in the top panel, shown after baseline subtraction and magnified for visibility.}
        \label{fig:D-Baseline-TE}
\end{figure}

Fig.~\ref{fig:D-Baseline-TE} shows a simulation of the Q-meter baseline with a spin-1 TE signal lineshape and Gaussian noise produced by MC sampling. In our simulations we match the baseline parameters, the baseline-to-signal scale, and the signal-to-noise ratio to realistic experimental values as closely as possible. The top panel shows how small the TE signal is relative to the baseline, with the inset emphasizing the two peaks of the Pake doublet. The bottom panel shows the same signal after baseline subtraction and magnification. The factor $C_E$ used in several vertical-axis labels denotes an experiment-dependent calibration scale; absolute vertical scales should therefore only be compared when the same calibration convention is used.

\subsection{Sources of Error}
\label{sec:errors}
The standard Liverpool Q-meter system is susceptible to several distinct sources of noise and systematic uncertainty that can influence the observed NMR signals. These contributions originate from power delivery and grounding, mechanical and electrical connectivity, the surrounding RF environment, and the intrinsic noise properties of the readout electronics. Collectively, they can introduce both stochastic noise and structured distortions into the measured spectra.

At the front end of the system, thermal noise generated by resistive elements in the resonant circuit, cabling, and preamplifier input stages contributes additive noise that is well approximated by Gaussian statistics. While this noise is often treated as spectrally white at the point of generation, it can become frequency dependent after passing through the resonant tank circuit, band-limiting filters, and frequency-dependent gain of the Q-meter chain, which can result in colored Gaussian noise in the measured spectrum.

Additional noise contributions arise from the RF environment and power infrastructure. Imperfect grounding, ground loops, and power-supply ripple can introduce coherent or quasi-coherent interference at specific frequencies, leading to narrowband noise or baseline modulation. External RF pickup, including environmental electromagnetic interference and cross-coupling from nearby RF systems, can further contaminate the signal with non-white and, in some cases, non-Gaussian components.

Mechanical factors such as loose connections, cable motion, and microphonics can modulate impedance and phase in the resonant circuit, producing low-frequency fluctuations and baseline instabilities that are correlated across frequency bins. These effects are not well described by ideal white-noise assumptions and can introduce structured distortions that persist across repeated measurements.

Finally, digitization and signal processing steps---including mixing, demodulation, filtering, and baseline subtraction---can reshape both the amplitude distribution and spectral content of the noise, further complicating its characterization. As a result, the noise observed in the final NMR spectra is generally a superposition of Gaussian and non-Gaussian components with frequency-dependent structure, rather than purely additive white Gaussian noise.

The more completely the environmental noise conditions are characterized, the more accurately they can be simulated. For the benchmark studies presented here, we focus on the dominant contributions that can alter the shape, stability, or interpretation of the resulting NMR signals, with the understanding that additional experiment-specific contributions can be added as needed. We include the following basic noise and baseline systematics:
\begin{itemize}
    \item \textbf{Gaussian Noise}: This is additive background noise that appears in every frequency bin of the recorded NMR data. This type of noise is suppressed, on average, as $1/\sqrt{N}$ when the number of accumulated NMR sweeps $N$ is increased.
    \item \textbf{Sinusoidal Noise}: Noisy power sources for the Q-meter system can produce sinusoidal interference that appears as either regular or irregular structured noise. This contribution is generally suppressed by averaging over a large number of sweeps (for example, $\sim 5000$) during baseline measurements, provided that the interference pattern is stable in time.
    \item \textbf{Shifts in Baseline}:  This effect typically manifests as a sudden shift in the tuning condition resulting in a skewed or tilted baseline. Such shifts can arise from several sources, including spontaneous changes in tuning due to loose $\lambda$/2 or phase-cable connections, temperature-induced drift within the system, or variations in capacitance between the $LC$ tank circuit and the phase detector. Any of these mechanisms can perturb the balance of the resonant circuit and introduce a persistent baseline distortion in the measured spectrum.
\end{itemize}

Standard extraction methods also introduce significant fitting errors that are not encoded in the simulations but are worth mentioning here. Time-dependent variations in the measured Q-curve or NMR signal, as well as imperfect baseline subtraction, can introduce a residual background in the spectrum. Traditional analysis methods attempt to remove this contribution by fitting and subtracting a low-order polynomial---typically third order, and in some cases second order---to the spectral wings. When this procedure is inadequate, the remaining background can distort the extracted NMR signal, leading to a systematic bias. This effect constitutes a baseline-induced fit error rather than statistical (Gaussian) noise.  When using Dulya-type fitting \cite{Dulya1997,Keller2017} off-center fits or poor baseline subtraction can result in false asymmetries in the two absorption intensities.  

Additionally, when calibration constants are used to map the integrated NMR area to polarization---most notably for the proton---the uncertainty in the extracted calibration constant obtained from the TE measurement can be dominated by large-scale noise relative to the signal amplitude. This effect can lead to a substantial relative uncertainty in the extracted polarization. Further contributions may arise if the spin system has not fully reached true thermal equilibrium at the time of calibration; however, this constitutes a systematic effect that is not addressed in the present work. The signal-to-noise limitations of TE measurements, along with the other error sources listed above, can be reliably modeled within the simulation framework employed here.

Specific DNN models can be trained using the simulated data to predict either the signal area or the polarization directly. The area-based approach is more general and can be applied to both spin-1/2 and spin-1 systems. In contrast, direct polarization prediction is best suited to spin-1 targets in materials with non-cubic symmetry, where each polarization value produces a distinct lineshape. The direct-polarization models therefore use the spin-1 lineshape theory as part of the training information to infer polarization under noisy or baseline-distorted conditions.

\section{Spin-1 Lineshape Theory}
\label{sec:theory}

\subsection{Polarization of Spin-1 Materials}
\label{sec:spin-1}

The spin-1 target materials considered in this work include ND$_3$ and deuterated butanol, for which the nuclei of interest are typically ${}^{2}\mathrm{H}$ or ${}^{14}\mathrm{N}$. In non-cubic crystalline environments, local electric-field gradients couple to the nuclear quadrupole moment and split the Zeeman transitions into two partially overlapping NMR absorption branches. This produces the characteristic spin-1 Pake-doublet lineshape used throughout this work for the direct polarization-regression studies. The detailed derivation of the spin-1 energy levels and transition frequencies is well established and is not repeated here; see Refs.~\cite{Abragam1961,Dulya1997,Keller2017,Keller2020} for the full treatment. Here we only summarize the analytical form required to generate the synthetic training spectra.

\begin{figure}[t]
    \centering
    \includegraphics[width=.45\textwidth]{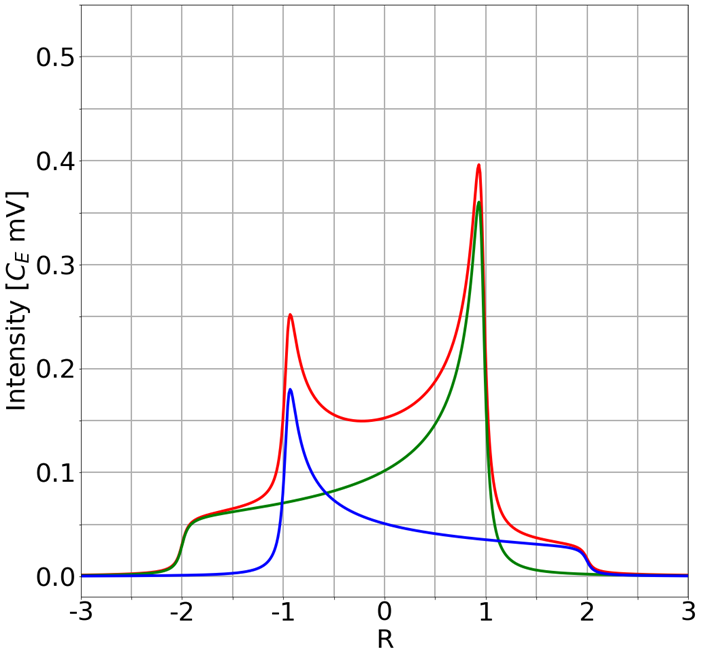}
    \caption{Simulated deuteron absorption lineshape showing $I_{+}$ (green) and $I_{-}$ (blue) solutions. The horizontal axis is the dimensionless detuning parameter $R$.}
    \label{fig:d-lineshape}
\end{figure}

Following the notation of Dulya et al.~\cite{Dulya1997}, the lineshape is expressed in terms of the reduced detuning variable
\begin{equation}
    R = \frac{\omega - \omega_0}{3\omega_Q},
\end{equation}
where $\omega$ is the probe frequency, $\omega_0$ is the Larmor frequency, and $\omega_Q$ is the quadrupole frequency. The two absorption branches are labeled by $\epsilon=\pm1$. For a given branch, the normalized analytical intensity is written as
\begin{equation}
\small
    \label{eq:icurve}
    \begin{aligned}
        I(R,\epsilon) = & \frac{1}{2\pi\mathscr{X}} \left[ 2\cos{\left(\frac{\alpha}{2}\right)} \left( \arctan\left(\frac{\mathscr{Y}^2-\mathscr{X}^2}{2\mathscr{Y}\mathscr{X}\sin{\left(\frac{\alpha}{2}\right)}}\right) + \frac{\pi}{2} \right) \right. \\
        & \left. + \sin{\left(\frac{\alpha}{2}\right)} \ln{\left(\frac{\mathscr{Y}^2+\mathscr{X}^2+2\mathscr{Y}\mathscr{X}\cos{\left(\frac{\alpha}{2}\right)}}{\mathscr{Y}^2+\mathscr{X}^2-2\mathscr{Y}\mathscr{X}\cos{\left(\frac{\alpha}{2}\right)}}\right)} \right],
    \end{aligned}
\end{equation}
with
\begin{align}
\label{eq:params}
\mathscr{X} &= \sqrt{A^2 + \left(1 - \epsilon R - \eta\cos 2\phi\right)^2}, \\
\mathscr{Y} &= \sqrt{3 - \eta\cos 2\phi}, \\
\cos\alpha &= \frac{1 - \epsilon R - \eta\cos 2\phi}{\mathscr{X}}.
\end{align}
Here $A$ is the dipolar broadening parameter, $\eta$ is the quadrupole asymmetry parameter, and $\phi$ is the azimuthal angle of the electric-field-gradient principal axis relative to the magnetic field. The full deuteron absorption signal is constructed from the two branches, $I(R,+1)$ and $I(R,-1)$, whose relative amplitudes encode the vector polarization. An example simulated lineshape is shown in Fig.~\ref{fig:d-lineshape}.

In the Monte Carlo training data, the ideal spin-1 target signal is therefore generated as a weighted sum of the two branches,
\begin{equation}
    S_{\mathrm{D}}(R)
    =
    C_E\left[a_{+}I(R,+1)+a_{-}I(R,-1)\right],
    \label{eq:spin1_training_signal}
\end{equation}
where $a_{+}$ and $a_{-}$ set the relative branch areas and $C_E$ is an experiment-dependent scale factor that absorbs the gain, coil filling factor, calibration normalization, and other setup-specific response factors. Thus, $C_E$ is not treated as a universal material constant, and spectra from different experiments or simulation configurations are not assumed to be directly comparable in absolute vertical scale. To generate a complete training event, this clean target response is superimposed on the simulated Q-meter baseline and noise described in Sec.~\ref{sec:qmeter}. In this way, Eq.~\ref{eq:icurve} supplies the physical spin-1 absorption profile, while the circuit model supplies the instrumental background and distortions that the network must learn to handle.

Equation~\ref{eq:spin1_training_signal} should be understood as the minimal single-site Pake-doublet model used for the benchmark studies in this work. Real spin-1 target spectra can require more elaborate material-dependent descriptions, including sums over chemically or crystallographically inequivalent sites, different quadrupole splittings, site-dependent broadening parameters, residual contaminant resonances, or small artifact structures introduced by the target material, irradiation history, or RF environment. These effects can be incorporated into the same simulation framework by replacing Eq.~\ref{eq:spin1_training_signal} with a weighted sum of additional constrained lineshape components. For the majority of this study, however, we use the minimal form in order to keep the training labels unambiguous and to isolate the performance of the machine-learning extraction on the dominant spin-1 lineshape together with realistic Q-meter baseline and noise distortions. 

\subsubsection*{Vector and Tensor Polarization}

For a spin-1 system with sublevel populations $n_{+1}$, $n_0$, and $n_{-1}$, the vector and tensor polarizations are defined as
\begin{align}
P_n &= \frac{n_{+1} - n_{-1}}{n}, \\
Q_n &= \frac{n - 3n_0}{n},
\end{align}
where
\begin{equation}
n \equiv n_{+1} + n_0 + n_{-1}.
\end{equation}
For normalized populations, $n=1$. Under Boltzmann-equilibrium conditions, the tensor polarization may be written in terms of the vector polarization as
\begin{equation}
    Q_n = 2 - \sqrt{4 - 3P_n^2}.
    \label{boltz}
\end{equation}

For vector-polarization extraction from the spin-1 Pake doublet, the relevant quantity is the ratio of the integrated branch intensities,
\begin{equation}
    r = \frac{\mathcal{I}_{+}}{\mathcal{I}_{-}},
\end{equation}
where $\mathcal{I}_{+}$ and $\mathcal{I}_{-}$ denote the areas of the two absorption branches. The vector polarization is then given by~\cite{Dulya1997}
\begin{equation}
    \label{eqn:polarization}
    P = \frac{r^2 - 1}{r^2 + r + 1}.
\end{equation}
This relation is particularly useful for simulation-based machine learning studies: a target vector polarization can be selected, the corresponding branch-area ratio $r$ can be computed, and the amplitudes $a_{+}$ and $a_{-}$ in Eq.~\ref{eq:spin1_training_signal} can then be chosen accordingly. More general spin-1 analyses can allow the two branch areas to vary independently~\cite{Keller2017}, which is required for enhanced tensor-polarization studies. For the benchmark vector-polarization studies presented here, we maintain the Boltzmann constraint in Eq.~\ref{boltz}.

\subsection{Materials with Cubic Symmetry}

Some spin-1 target materials, such as ${}^6$LiD, place the spin-1 nucleus in an approximately cubic crystal field. In this limit, the local EFG is strongly suppressed, the quadrupolar splitting becomes unresolved, and the spin-1 NMR response appears as a single proton-like resonance rather than a Pake doublet. This single-line response is the relevant case for the area-focused models and is sufficiently represented by a Voigt profile, i.e.\ a convolution of Gaussian and Lorentzian contributions,
\begin{equation}
    f(x;\sigma,\gamma_L) \equiv \int_{-\infty}^{\infty} G(x';\sigma)\,L(x-x';\gamma_L)\,dx',
\end{equation}
where $G$ and $L$ denote the Gaussian and Lorentzian distributions, respectively~\cite{Abragam1961}. The Gaussian width $\sigma$ represents static or quasi-static inhomogeneous broadening from dipolar fields, bulk-susceptibility variations, and related field-distribution effects, while the Lorentzian width $\gamma_L$ represents homogeneous broadening from spin--spin relaxation and time-dependent local-field fluctuations~\cite{abragam_principles_1986}. Under strong microwave irradiation for dynamic nuclear polarization, power broadening can further increase the Lorentzian contribution.

Thus the simulation suite uses two primary signal models: the Pake doublet for non-cubic spin-1 systems where lineshape asymmetry carries vector-polarization information, and the Voigt profile for proton-like single-line signals where the integrated area is the primary observable. More detailed derivations of these lineshapes and their polarized-target applications are given in Refs.~\cite{Dulya1997,Keller2017,Abragam1961}.

\begin{figure}[t]
    \centering
    \includegraphics[width=1.0\linewidth]{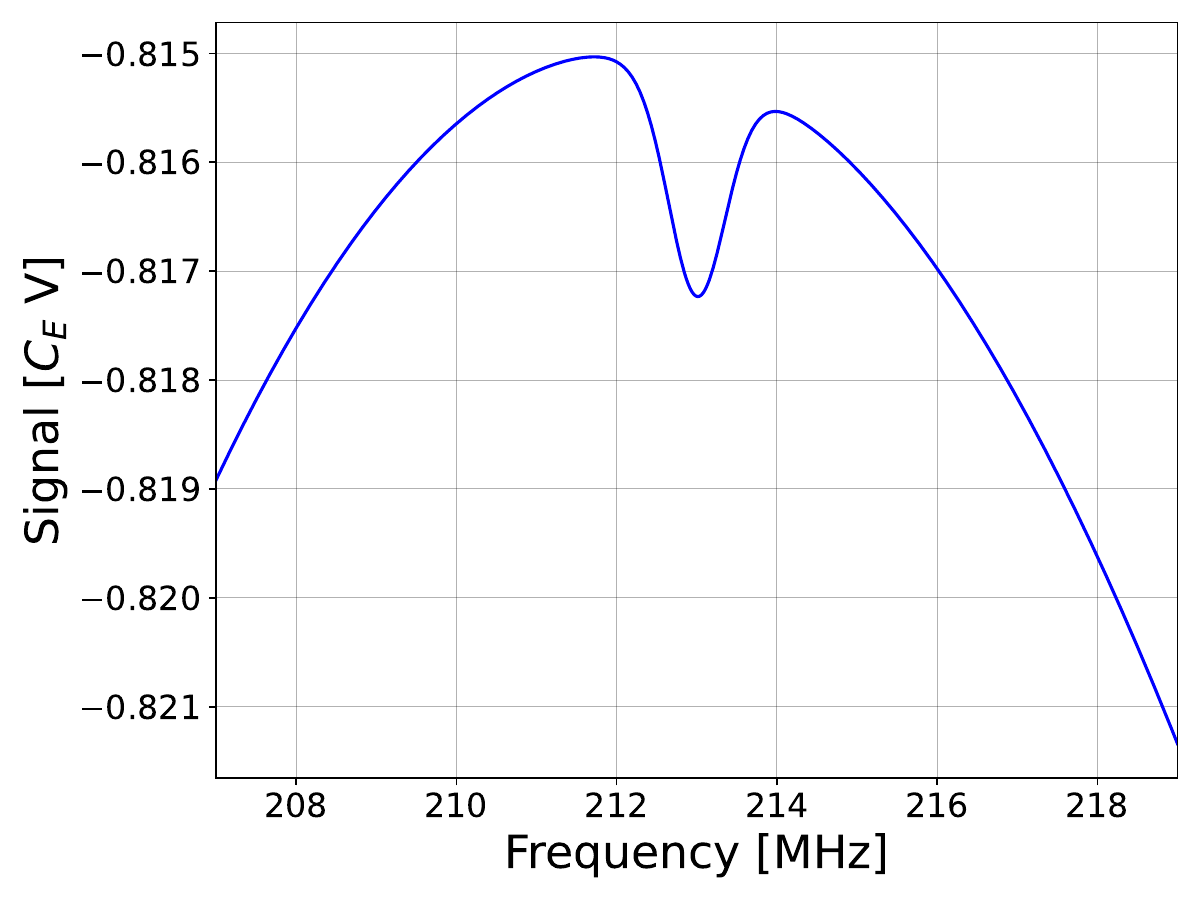}
    \caption{Simulated proton NMR signal demonstrating a Voigt-type lineshape. Parameters for Voigt profile simulation were extracted from experimental fits.}
    \label{fig:proton_sample}
\end{figure}

\section{Historical Extractions and Error}
\label{sec:historical}
The error contribution explored in this work is the fitting uncertainty associated with the CW--NMR polarization extraction. In principle, modern AI-based analysis methods should be capable of significantly reducing this source of uncertainty under appropriate experimental conditions. We therefore first review the level of fitting uncertainty achieved by traditional extraction techniques and the extent to which these methods have been able to directly minimize this contribution to the total polarization error budget.

For enhanced deuteron signals, the original work of Dulya \textit{et al.} quoted a conservative relative uncertainty of approximately \(3\%\) from the fitting procedure alone, with a total relative polarization uncertainty of about \(5\%\)~\cite{Dulya1997}. However, this fitting method is not normally applied at the thermal-equilibrium (TE) polarization scale.

The direct extraction of deuteron polarization from TE CW-NMR signals is intrinsically challenging because the expected polarization is extremely small. For a spin--1 deuteron target under typical TE measurement conditions, for example \(5~\mathrm{T}\) and \(2~\mathrm{K}\), the polarization is only of order \(0.05\%\). Consequently, the two components of the Pake-doublet lineshape differ only weakly in both integrated area and peak amplitude. This makes a direct lineshape-based extraction of the TE polarization highly sensitive to noise, baseline treatment, and the assumed fitting model. For this reason, deuteron analyses have more commonly relied on direct fitting of the enhanced DNP signal, rather than on full lineshape reconstruction at the much smaller TE scale.

For TE-scale deuteron measurements, the calibration procedure has historically relied on the integrated-area method, which is also the standard approach for proton TE and enhanced-polarization measurements~\cite{Adams1999,Adeva1994,Dhawan1996,Reicherz2024}. In this method, the TE signal is averaged over many sweeps, the baseline is subtracted, and the remaining absorptive signal area is obtained by numerical integration, typically through a Riemann-sum or bin-integration procedure. The extracted area is then compared with the known TE polarization. Using this approach, Adams \textit{et al.} reported one of the smallest quoted fitting uncertainties for a TE-scale deuteron signal, namely \(0.83\%\), together with a final polarization uncertainty of \(2.0\%\)~\cite{Adams1999}.

The most relevant prior work on Dulya-type deuteron lineshape fitting showed that a full lineshape fit can be applied at the TE scale only after very substantial averaging.  In Ref.~\cite{Dulya1997}, the authors constructed a high-statistics ``super-TE'' deuteron signal by averaging many individual TE measurements.  This was necessary because ordinary TE spectra did not provide sufficient signal-to-noise ratio to reliably determine the detailed lineshape parameters.  The fitted super-TE signal was then used primarily to study and constrain instrumental false-asymmetry effects in the Q-meter response, rather than to replace the standard TE area calibration.

This distinction is important for the interpretation of the achievable error.  The Dulya-type fit contains both physical lineshape parameters and instrumental/background parameters.  At large polarization, the relative asymmetry of the two deuteron transitions is large enough that the fit can constrain the polarization-sensitive part of the model.  At TE, however, the true asymmetry is very close to unity, so small residual baseline distortions, phase-mixing effects, or Q-meter false asymmetries can mimic a polarization-dependent distortion of the lineshape.  Thus, a fully unconstrained fit to a weak TE deuteron signal is under-determined in practice: the fit can trade off polarization, background, phase, and false-asymmetry parameters while producing visually acceptable residuals.

For proton-type signals analyzed with the area method, the extraction uncertainty includes the numerical integration or fit uncertainty of the TE signal, the baseline or polynomial-background subtraction, and possible response-function drifts of the Q-meter circuit.  
Several polarized-proton target analyses provide useful benchmarks for this contribution.  In the SMC proton measurement, the polarization calibration constant was obtained from TE signals at approximately 1~K, and the accuracy of the TE calibration signal contributed $\Delta P/P = 1.1\%$ to the polarization error budget~\cite{Adams1997SMCProtonG1}.  The underlying high-precision SMC target-polarization study reported a final relative polarization accuracy of about $3\%$, with the principal uncertainty groups arising from the calibration temperature, TE NMR signal, enhanced-polarization NMR signal, and low-frequency gain ratio~\cite{Kramer1995HighPrecision}.  For polarized ammonia, the SMC ammonia study reported routine proton polarizations of approximately $\pm(90\pm2.5)\%$ and provides an important solid-${}^{14}\mathrm{NH}_3$ benchmark for TE-based proton calibration in a large target system~\cite{Adeva1998AmmoniaEST}.

A more explicit decomposition of the TE extraction terms was given in the JLab E08-007 analysis of irradiated ${}^{14}\mathrm{NH}_3$ targets.  In that study, the TE-area uncertainty from the Riemann-sum integration procedure was bounded by the relative error of 1.61\%,
while the polynomial-background or fit contribution was bounded by a relative error of 0.75\%. The same work reported a total relative target-polarization uncertainty of less than or equal to $3.9\%$ after including the other calibration and instrumental contributions~\cite{KELLER2013133}.  

The smallest directly relevant values are therefore not a single universal number: fit/background components can be below $1\%$, TE signal-area or TE calibration-signal contributions are commonly near $1$--$2\%$, and the complete target-polarization uncertainty is generally much larger once thermometry, Q-meter response, circuit nonlinearity, and other systematic effects are included.  This distinction is important for the present AI-based study, since the objective is not to replace the full TE calibration procedure, but to reduce the signal-extraction component of the CW--NMR polarization error budget.

\section{Methodology for Extraction of Polarization}
\label{sec:methodology}

In this section, we review traditional polarization extraction techniques, including thermal-equilibrium measurements and lineshape fitting.  We then introduce the DNN approaches.
    
\subsection{Thermal Equilibrium Technique}
The area-based calibration method relies on a thermal-equilibrium (TE) measurement to establish an absolute normalization relating the integrated NMR signal area to the polarization (vector polarization in the case of spin-1 targets). When the lattice and target material are allowed to reach true thermal equilibrium, the resulting polarization is uniquely determined by the spin quantum number and follows the well-defined Boltzmann distributions for spin-1/2 and spin-1 systems:
\begin{equation}
P_{\mathrm{TE}}^{S=\frac{1}{2}}
=
\tanh\!\left(\frac{g\mu B}{2kT}\right).
\end{equation}

\begin{equation}
P_{\mathrm{TE}}^{S=1}
=
\frac{4\tanh\!\left(\frac{g\mu B}{2kT}\right)}
{3+\tanh^2\!\left(\frac{g\mu B}{2kT}\right)}.
\end{equation}

where $B$ is the external magnetic field, $\mu$ the magnetic moment in the external field of strength $B$, $k$ is the Boltzmann constant, $g$ is the g-factor of the specimen, and $T$ is the temperature of the system~\cite{Crabb1997}.

For a typical calibration temperature of $T = 1.5~\mathrm{K}$, NH$_3$ yields a reference polarization of
\begin{equation}
P_{\mathrm{TE}}^{S=\frac{1}{2}} \approx 0.341\%.
\end{equation}
For ND$_3$,
\begin{equation}
P_{\mathrm{TE}}^{S=1} \approx 0.0697\%.
\end{equation}
Once this reference value is established, any non-equilibrium polarization can be determined by integrating the measured absorption signal \cite{Dulya1997},
\begin{equation}
P = C \int \frac{d\omega\, S(\omega)}{\omega},
\end{equation}
where \(S(\omega)\) denotes the extracted NMR absorption signal as a function of angular frequency. In the context of a Q-meter-based CW-NMR measurement, this signal is defined as
\begin{equation}
S(\omega) \equiv \Re\!\left\{ V(\omega,\chi) - V(\omega,0) \right\}
\;\propto\; \chi^{\prime\prime}(\omega),
\end{equation}
with \(V(\omega,\chi)\) the measured complex voltage across the resonant circuit in the presence of nuclear susceptibility \(\chi(\omega)\), and \(V(\omega,0)\) the corresponding background voltage measured in the absence of the NMR response. The quantity \(\chi^{\prime\prime}(\omega)\) is the absorptive component of the complex magnetic susceptibility, which carries the polarization-dependent signal. The proportionality constant \(C\) is a calibration factor determined from the thermal-equilibrium measurement and provides the absolute normalization relating the integrated signal area to the physical polarization.

The constant $C$ is obtained directly from a TE calibration measurement via
\begin{equation}
C =
\frac{\displaystyle P_{\text{TE}}}
{\displaystyle \int\frac{d\omega\ S(\omega)}{\omega}},
\end{equation}
where $P_{\text{TE}}$ is the TE polarization of either a spin-1/2 or spin-1 sample and $T$ is taken as the experimentally measured lattice temperature \cite{Crabb1997}.

The thermal-equilibrium (TE) calibration method can introduce substantial uncertainty, resulting in a total relative error of 3--5\% for proton polarization under realistic experimental conditions \cite{Keller2017}. For spin-1 targets, the corresponding uncertainty is generally larger; moreover, even a vector-polarization uncertainty of this magnitude propagates to a relative uncertainty exceeding 7\% in the extracted tensor polarization~\cite{Keller2020,KELLER2013133,Kielhorn,MeyerSchilling} when using the standard Boltzmann relationship in Eq.~\ref{boltz}. Such deviations commonly arise from a limited number of TE calibration points or from insufficient time for the target material to reach true thermal equilibrium. In practice, when relying on the TE method, these sources of uncertainty are often unavoidable due to beam-time constraints and operational priorities. By contrast, the uncertainty component arising purely from spectral fitting and area integration is typically at the 2--3\% level~\cite{KELLER2013133}, representing the dominant avenue where methodological improvements can be achieved.

\subsection{Dulya (and Dulya-Like) Fitting}

For spin-1 targets with non-cubic symmetry, lineshape fitting methods such as the Dulya method~\cite{Dulya1997} and Dulya-like fits~\cite{Keller2017} use the analytical lineshape in Eq.~\ref{eq:icurve} to fit directly to the NMR signal. In a standard workflow, the baseline is first subtracted, a low-order polynomial is fit to any residual background, and that residual contribution is removed in a manner similar to the area approach. Fits can also be performed without explicit baseline subtraction when the background terms are included in the fit model. A $\chi^2$ minimization is then used to extract the intensity ratio of $I_+$ and $I_-$. The fitted absorption-line heights, or in related methods the fitted absorption-line areas~\cite{Keller2017}, determine the ratio $r$ and therefore the vector polarization via Eq.~\ref{eqn:polarization}. For enhanced spin-1 signals, this lineshape-based approach generally has a smaller relative extraction error than a TE-area calibration because the enhanced signal has much higher signal-to-noise than a TE-scale signal and does not require waiting for the target to thermalize.
  
\subsection{Artificial Neural Networks (ANN)}
Neural networks are trained by repeatedly presenting labeled examples---each consisting of an input and its corresponding output---and adjusting internal parameters to strengthen the mapping between them. Through this iterative optimization, the model encodes a probability-weighted representation of the underlying functional relationship within its parameters. While the literature on artificial neural networks (ANNs) is vast, we provide only a concise overview here.

A neural network may be viewed as a composition of nonlinear functions---commonly referred to as \textit{activation functions}---denoted by $\phi_i$. For a given network architecture, one constructs a sequence of transformations
\[
\phi_1,\, \phi_2,\, \ldots,\, \phi_L
\]
whose composition aims to approximate a target function $\hat{f}$ according to an optimization criterion, typically defined by a loss function \cite{hastie01statisticallearning}. The Universal Approximation Theorem guarantees that, under mild assumptions, sufficiently large networks can approximate any continuous function $\hat{f}$ to arbitrary precision $\epsilon$ \cite{Hornik1989}. However, the theorem is non-constructive: it does not specify the required architecture, depth, width, or choice of nonlinearities.

In practice, network accuracy depends sensitively on \textit{hyperparameters}---including depth, width, learning rate, regularization, and activation type---which must be tuned for the problem at hand. Training proceeds by forward-propagating an input through the network to produce a prediction, comparing that prediction to the target output via the loss function, and then updating the network weights through backpropagation. Repeating this process over many examples iteratively reduces the loss, yielding a model whose outputs converge toward the desired target values.

\begin{figure}[h!]
    \centering
    \includegraphics[width=0.5\textwidth]{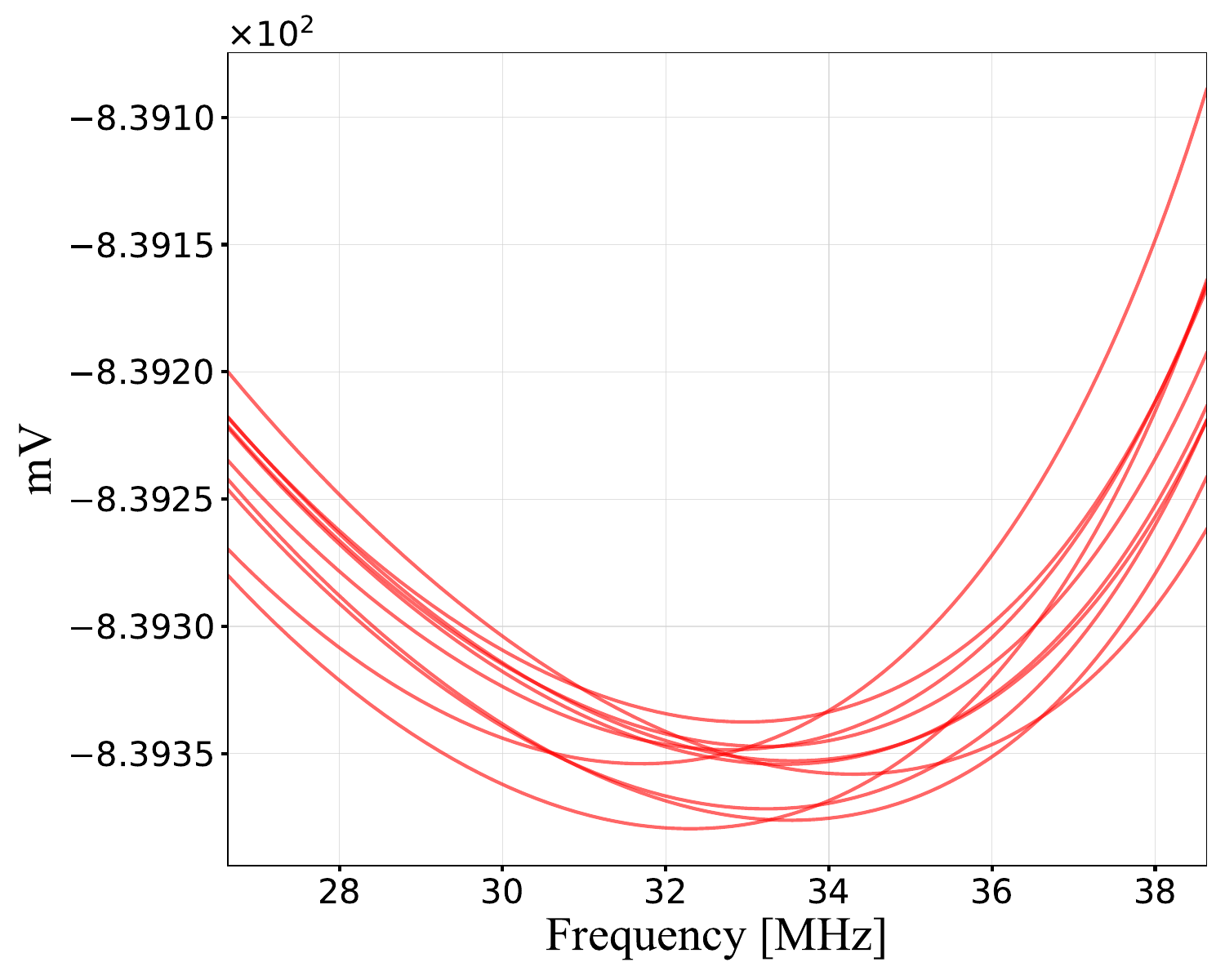}
    \caption{Examples of applied variations in the simulated baseline in diode mode.}
    \label{fig:Shifting}
\end{figure}

Inspired by data augmentation methods commonly used in computer vision \cite{kumar2023imagedataaugmentationapproaches,shorten2019survey}, we introduced controlled transformations to the simulated baseline signals in diode mode of the Q-meter, shown in Fig.~\ref{fig:Shifting}. These baseline shifts arise from small perturbations in three physical parameters of the Q-curve model: the applied voltage $U$, the adjustable capacitance $C_{\mathrm{knob}}$, and the phase $\varphi$. Let $x \in \mathcal{D}$ denote a clean, simulated signal prior to augmentation. We define an augmentation operator
\begin{equation}
    T_{\theta}(x) = x + g(U + \Delta U,\, C_{\mathrm{knob}} + \Delta C,\, \varphi + \Delta\varphi),
\end{equation}
where $\theta = (\Delta U, \Delta C, \Delta\varphi)$ represents small, physically realistic perturbations sampled from bounded distributions, and $g(\cdot)$ represents the baseline-shape generating function of the circuit model. The augmented dataset is then
\begin{equation}
    \widetilde{\mathcal{D}} = \{T_{\theta}(x) \mid x \in \mathcal{D},\ \theta \sim P(\theta)\}.
\end{equation}
As an example, for a deuteron setup at $5~\mathrm{T}$ with the cable length set to one $\lambda/2$, $U$ is varied between $0.001$ and $1$~V, while $C_{\mathrm{knob}}$ is varied between $10$ and $100$~pF (with optional change of fixed capacitance), and phase is varied at $\pm$20\% from its optimal tune ($2\pi$).  We note that phase changes affect the signal characteristics as well, which are a critical part of the training to make the model more generally robust to unintended tune shifts.

The motivation for this augmentation strategy is two-fold. First, it increases the effective sample size $|\widetilde{\mathcal{D}}|$, which is known to reduce overfitting and improve generalization bounds through variance reduction in empirical risk minimization \cite{vapnik1998statistical,zhang2021survey}. Second, it expands the functional variability of the data distribution, thereby increasing the diversity of the hypothesis space the model must approximate. Formally, if $\mathcal{F}$ denotes the class of real-valued measurable functions capable of mapping signals to polarization values, i.e.,
\begin{equation}
    \mathcal{F} : \widetilde{\mathcal{D}} \rightarrow \mathbb{R},
\end{equation}
then augmentation can be viewed as broadening the support of the empirical data distribution toward the true underlying generative distribution encountered in real measurements. This reduces covariate shift and improves robustness under domain variability \cite{goodfellow2016deep}.

The simulated lineshape of the signal is characterized by extracting the lineshape parameters using the standard fitting procedures \cite{Keller2017,Dulya1997} to experimental data.  The parameterized simulated signal is added to the simulated Q-curve, along with various types of noise to make high-quality training and test data.

The applied controlled transformations are intended to replicate realistic environmental effects in the Q-curve and signal, accounting for phase shifts that produce lineshape distortions associated with the dispersive response of the signal, in addition to systematic modifications to the spectral baseline.  Additionally, the simulated noise described in Sec.~\ref{sec:errors} is injected at various scales during training to mimic realistic scenarios and further generalize lineshape features. 

In short, by incorporating realistic noise and physical perturbations of baseline and tune parameters, our augmentation approach increases the representational richness of the dataset and enhances the model's capacity to generalize to unseen experimental signals rather than memorizing parametrically constrained simulated profiles.

We trained four separate neural network models: 
\begin{enumerate}
    \item a polarization model for the 2\%--60\% polarization region (hereafter referred to as the \textit{``high-polarization model''}),
    \item a polarization model for the TE--2\% polarization region (hereafter referred to as the \textit{``low-polarization model''}),
    \item an area-prediction model (\textit{``area model''}) targeting the TE--100\% range, with particular emphasis near the TE region for the proton,
    \item a denoising autoencoder (DAE) used to filter noise from signal data.
\end{enumerate}

For the high-polarization model, the lower bound of $2\%$ is chosen because the lineshape continues to exhibit a clearly quantifiable asymmetry between the two absorption lines, which remains discernible even in the presence of moderate noise. The upper bound of $60\%$ reflects the typical polarization range of our case study; however, the model can be straightforwardly extended to $100\%$ polarization by incorporating additional training data at higher polarization values.  Overall performance is discussed in Sec.~\ref{sec:results}.

A separate low-polarization model is required because polarization extraction in this regime constitutes a fundamentally different inference problem. At polarization levels near the TE scale, the signal becomes strongly noise-dominated and the signal-to-baseline ratio can become extremely small. These conditions suppress the relevant spectral features and can lead to poor feature learning in a general-purpose model. This limitation motivated careful training-data generation that mimics the settings of the Q-meter and preserves a consistent baseline scale and relative signal magnitude. In this configuration, the model can use the magnified signal scale relative to the baseline as an important training feature, at the cost of broader generalization. The resulting inference is effectively based on both signal area and lineshape information, leading to the improved results discussed in Sec.~\ref{sec:results}.

The area model is designed to work with any signal lineshape and is generally required for proton-type signals and other spectra for which a specific lineshape is not needed for accurate inference. Since the low-polarization model targets the TE-scale region of Pake-doublet spin-1 signals, the area model focuses on proton-scale TE signals. This makes the task simpler: the model primarily needs a high-quality extraction of signal areas corresponding to approximately $0.3\%$ polarization.

Finally, the DAE serves a distinct role from the other models: its primary purpose is to suppress noise in the NMR signal, rather than to directly infer polarization values.  The cleaned signal can then be better analyzed with other tools to evaluate the spectra or polarization.

\subsection{Analysis Metrics}
\label{sec:metrics}
Extraction error, or fit error, is defined here as the uncertainty introduced by the polarization-extraction procedure itself. This may include statistical uncertainty from numerical integration, fit uncertainty from the chosen signal model, bias inherent to the extraction method, and systematic effects associated with baseline subtraction, signal distortion, or an off-centered NMR lineshape. These contributions can be studied collectively by constructing simulated signals that reproduce the relevant experimental conditions, including the expected signal scale, noise level, baseline structure, and possible lineshape distortions. Repeated pseudo-experiments can then be performed in which these effects are varied systematically and the polarization is extracted from signals with known input areas or known true polarizations. This provides a direct and controlled benchmark of the extraction procedure.

Because the true input polarization is known in simulation, the performance of the extraction method can be quantified directly from the residual distribution between the true and extracted polarizations. For each test sample, we define the pointwise polarization residual as
\begin{equation}
\Delta P_n = P^{\mathrm{true}}_n - P^{\mathrm{pred}}_n,
\end{equation}
where \(P^{\mathrm{true}}_n\) is the known simulated polarization and \(P^{\mathrm{pred}}_n\) is the corresponding model prediction. A positive residual therefore indicates that the model underpredicts the true polarization, while a negative residual indicates overprediction. For the area model, the same definitions are applied to the area residual \(\Delta A_n=A^{\mathrm{true}}_n-A^{\mathrm{pred}}_n\).

The bias of the extraction method is characterized by the mean residual,
\begin{equation}
\mu = \frac{1}{N}\sum_{n=1}^{N}\Delta P_n,
\end{equation}
where \(N\) is the total number of test samples. The precision is characterized by the standard deviation of the residual distribution,
\begin{equation}
\sigma =
\sqrt{
\frac{1}{N}\sum_{n=1}^{N}
\left(\Delta P_n-\mu\right)^2
}.
\end{equation}
Thus \(\mu\) measures the systematic offset of the extraction procedure, while \(\sigma\) measures the random or sample-to-sample spread about that offset. Unless explicitly stated otherwise, residual means and widths reported in the tables are absolute residuals: polarization residuals are given in units of polarization percent, and area residuals are given in units of integrated signal area.

If the bias is not corrected, the total extraction error in absolute units is the root-mean-square residual,
\begin{equation}
\epsilon_{\mathrm{tot}}
=
\sqrt{
\frac{1}{N}\sum_{n=1}^{N}\Delta P_n^2
}
=
\sqrt{\mu^2+\sigma^2}.
\end{equation}
For a fixed-polarization study, or for a narrow polarization bin with representative true polarization \(P_{\mathrm{true}}\), the total relative extraction contribution is then
\begin{equation}
\epsilon_{\mathrm{rel}} = \frac{\epsilon_{\mathrm{tot}}}{P_{\mathrm{true}}}.
\end{equation}
This relative quantity, usually reported as a percentage, is the most direct comparison to conventional experimental polarization-error budgets. The absolute residuals \(\mu\) and \(\sigma\) are retained because they remain well defined near the TE scale and avoid instabilities associated with dividing each event by a very small polarization.

Noise levels for each event were quantified using the signal-to-noise ratio (SNR), defined as
\begin{equation}
    \mathrm{SNR} = \frac{\max(|\text{Signal}|)}{\max(|\text{Noise}|)}.
    \label{eqn:SNR}
\end{equation}
An SNR value of 1 indicates that the noise amplitude is equal in magnitude to the signal amplitude. Values significantly greater than 1 correspond to comparatively low-noise events, while values much less than 1 indicate highly noise-dominated events. In this context, the ``signal'' refers specifically to the ND$_3$ lineshape, excluding the baseline contribution, as shown in Fig.~\ref{fig:D-Baseline-TE}. Because the same absolute noise scale produces different event-level SNR values for different polarization ranges and signal models, SNR values quoted in different studies are not expected to be identical. When ``low'' and ``high'' noise labels are used below, they refer to the injected noise-amplitude scale for that study; the event-level SNR is computed from Eq.~\ref{eqn:SNR}.

\section{Model Construction}
\label{sec:architecture}
To address the distinct inference tasks involved in polarization extraction, signal-area determination, and noise suppression, we employ a suite of specialized neural-network architectures. Convolutional neural networks (CNNs) are evaluated in both the high- and low-polarization regimes, where localized spectral features and lineshape asymmetries contain the dominant physical information. A simpler multilayer perceptron (MLP) architecture is also evaluated for both polarization regimes and for signal-area determination, where global amplitude information is sufficient and explicit spatial feature extraction is not required. For each task, the final architecture is selected based on validation performance. In addition, a denoising autoencoder (DAE) is used to suppress noise and recover clean spectral structure before downstream analysis. The following subsections describe the architecture, training strategy, and intended role of each model in detail.

\subsection{Polarization Models}
We address the polarization- and signal-extraction tasks using both multilayer perceptron (MLP) and convolutional neural network (CNN) architectures \cite{inproceedings}. Both are feed-forward neural-network models trained using gradient-based optimization: the MLP provides a simple fully connected baseline, whereas the CNN employs convolutional kernels to exploit local structure in the input spectrum. CNNs are widely used in modern machine learning, particularly in computer vision, image analysis, and spatiotemporal signal processing, because of their parameter efficiency, translation-equivariant structure, and strong inductive bias toward local feature extraction.

In the context of CW-NMR spectra, these properties are especially well matched to the underlying physics of the signal. The relevant information is encoded in localized spectral features---such as peak structure, relative lobe asymmetry, linewidth, and baseline curvature---that are largely invariant under small frequency shifts and scale changes induced by tuning drift, phase offsets, or baseline distortions. Convolutional kernels naturally capture such local correlations while remaining insensitive to their absolute position in frequency space, making CNNs well suited for robust polarization inference under non-ideal experimental conditions.

Furthermore, the implicit regularization introduced by convolutional weight sharing substantially reduces the number of free trainable parameters relative to fully connected architectures. This improves generalization, reduces overfitting, and mitigates training instabilities such as vanishing or exploding gradients \cite{Venkatesan}. Taken together, these characteristics make CNN-based architectures a natural and physically motivated choice for extracting polarization and signal observables from noisy, baseline-distorted NMR spectra.

\begin{figure}
    \centering
    \includegraphics[width=\linewidth]{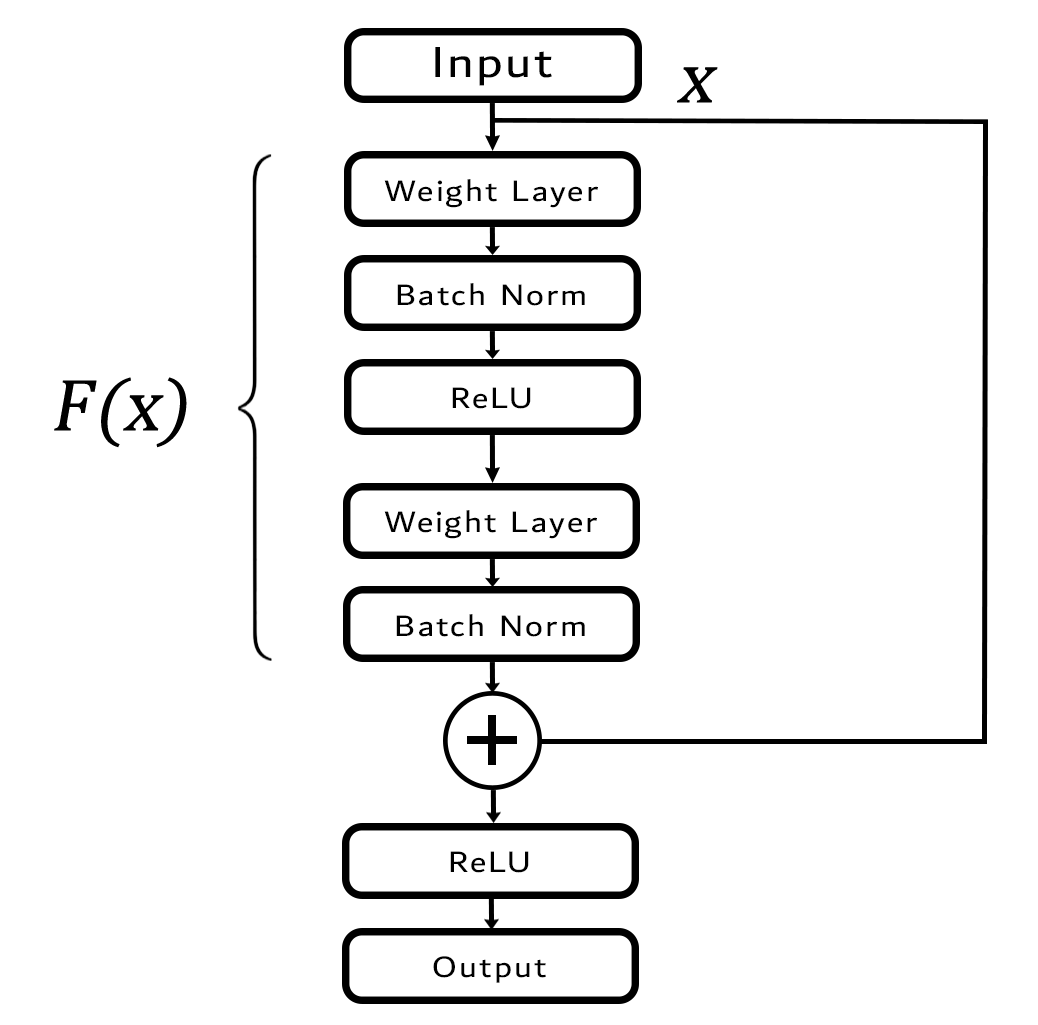}
    \caption{Example architecture of Residual Block (ResBlock).}
    \label{fig:resblock}
\end{figure}

A core architectural component of our model is the use of residual connections. Residual connections (or residual blocks) enable direct signal pathways within deep networks, improving gradient flow during backpropagation and allowing substantially deeper architectures to be trained without degradation \cite{he2015deepresiduallearningimage}. A residual block takes an input $x$ and outputs
\begin{equation}
    \operatorname{ResBlock}(x) = F(x) + x,
\end{equation}
where $F(\cdot)$ is a learnable nonlinear transformation (typically a composition of convolution, normalization, and activation layers). In the original formulation of \cite{he2015deepresiduallearningimage}, the underlying mapping $H(x)$ is expressed as $H(x) = F(x) + x$, where $F(x) := H(x) - x$ is known as the \textit{residual function}. An example figure of a ResBlock is shown in Fig.~\ref{fig:resblock}. The use of residual connections is now ubiquitous in modern deep learning due to demonstrated improvements in stability, convergence, and predictive performance across domains.

We also incorporated a multi-scale convolution block, also known as an \textit{Inception Block}~\cite{szegedy2014goingdeeperconvolutions}, designed to (1) leverage convolutions across multiple scales of spatial information and (2) reduce model complexity while increasing performance. The basic premise, as described in \cite{szegedy2014goingdeeperconvolutions}, involves applying several independent convolutions to the same input and then concatenating their outputs. As a preliminary step, we used one such block after the initial layer. More sophisticated models could include several, or even dozens, of these blocks within a single architecture.

Finally, prior to global pooling, we incorporate a \textit{squeeze-and-excitation (SE) block}, inspired by the framework introduced in Squeeze-and-Excitation Networks~\cite{hu2019squeezeandexcitationnetworks}. Consider a transformation $\mathbf{T}$ that maps an input $\mathbf{X}$ to a set of feature maps $\mathbf{U} \in \mathbb{R}^{H \times W \times C}$, where $H$, $W$, and $C$ denote the spatial dimensions and number of channels produced by a convolutional operation. An SE block adaptively recalibrates these feature maps by explicitly modeling channel-wise interdependencies, thereby enhancing representational capacity and overall model performance.

\begin{figure}
    \centering
    \includegraphics[width=0.62\linewidth]{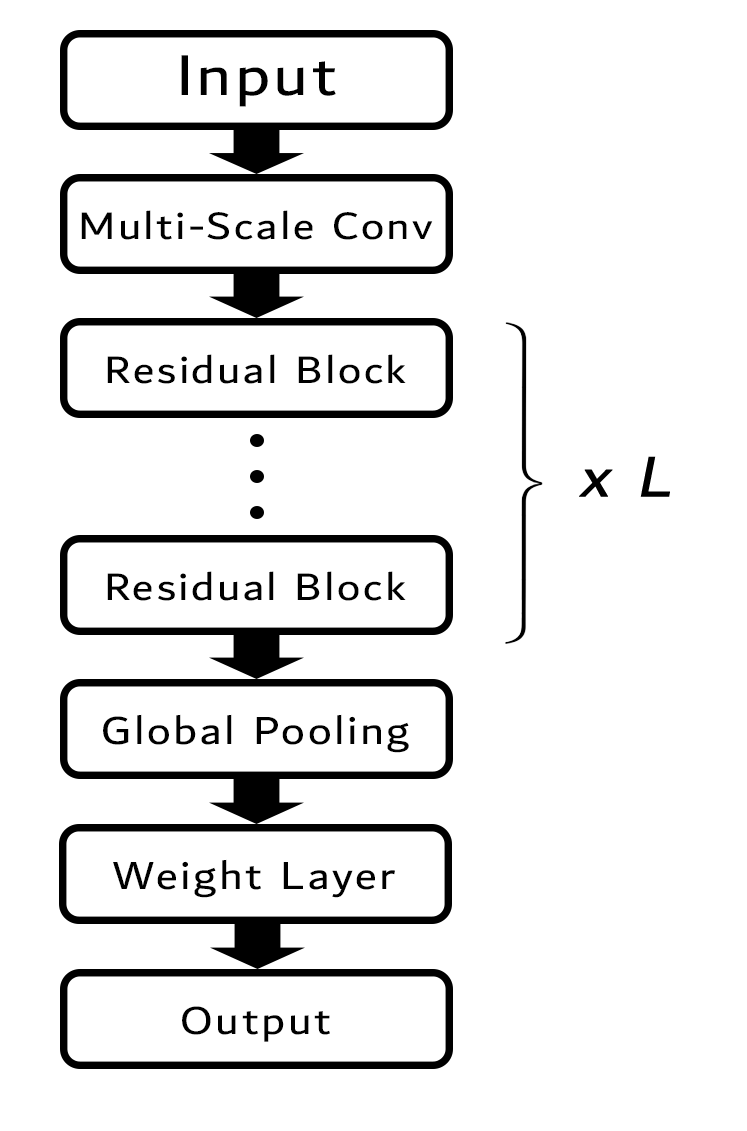}
    \caption{Abstract architecture diagram of the CNN-based polarization model. Hyperparameters varied between the high- and low-polarization implementations.}
    \label{fig:architecture}
\end{figure}

The \emph{squeeze} operation performs global average pooling across the spatial dimensions $H \times W$, generating a compact channel descriptor that captures the global distribution of feature responses. This is followed by an \emph{excitation} operation, which maps the channel descriptor to a set of per-channel modulation weights. These weights are then applied to $\mathbf{U}$ via channel-wise scaling, producing the output of the SE block, which is subsequently passed to downstream layers of the network. In practice, this self-gating mechanism is akin to a computationally efficient attention mechanism, which weights relevant information within feature maps~\cite{hu2019squeezeandexcitationnetworks,vaswani2023attentionneed}. Fig.~\ref{fig:architecture} shows a high-level diagram of the CNN-based architecture used to build the polarization models.

Although the high- and low-polarization models share an identical architectural layout and block structure---including the use of specialized Inception-style modules and SE blocks---their optimization procedures were performed independently. As a result, differences in the optimized filter depths, channel widths, and regularization parameters led to distinct effective parameterizations and, consequently, different numbers of free trainable parameters in each model.

\begin{table}[h!]
\centering
\begin{tabular}{lc}
\hline
\textbf{Hyperparameter} & \textbf{Value} \\ \hline
\# Layers           & 2 \\
\# Nodes           & 256 \\
$\alpha$ (Learning Rate)  & $3.0\times10^{-4}$ \\
Activation                & ReLU \\
Batch Size                & 256 \\
$\epsilon$                & $1\times10^{-8}$ \\
$\beta_1$                 & 0.99 \\
$\beta_2$                 & 0.999 \\
clipnorm                  & 1 \\\hline
\end{tabular}
\caption{Tuned hyperparameters for the MLP model.}
\label{tab:ff-parameters}
\end{table}

\begin{table}[h!]
\centering
\begin{tabular}{lc}
\hline
\textbf{Hyperparameter} & \textbf{Value} \\ \hline
\# of Residual Blocks     & 4 \\
Filters per Layer         & 64 \\
$\alpha$ (Learning Rate)  & 0.0081694 \\
Batch Size                & 256 \\
$\epsilon$                & $3.6438\times10^{-5}$ \\
$\beta_1$                 & 0.2936 \\
$\beta_2$                 & 0.2334 \\
clipnorm                  & 4.9805 \\ \hline
\end{tabular}
\caption{Tuned hyperparameters for the CNN model where Optuna was used to optimize hyperparameters.}
\label{tab:polarization_parameters}
\end{table}

Tables~\ref{tab:ff-parameters} and~\ref{tab:polarization_parameters} summarize the optimized hyperparameter configurations (the selection of these choices is discussed in Sec.~\ref{sec:training}). The hyperparameters $\epsilon$, $\beta_1$, $\beta_2$, $\alpha$, and \texttt{clipnorm} correspond to components of the \texttt{AdamW} optimizer \cite{loshchilov2019decoupledweightdecayregularization}, a decoupled variation of \texttt{Adam} that separates $L_2$-based weight decay regularization from gradient-based updates. Here, $\epsilon$ is a small constant added to the denominator for numerical stability, while $\beta_1$ and $\beta_2$ are exponential decay coefficients for the first- and second-order moment estimates of the gradient, respectively. With initial learning rate $\alpha$ and weight-decay parameter $\lambda$, the parameter update rule is
\begin{equation}
    \label{eqn:adamWupdate}
    \theta_{t+1} = \theta_t - \alpha \left(\frac{\hat{m}_t}{\sqrt{\hat{v}_t} + \epsilon} + \lambda \theta_t\right),
\end{equation}
where $\hat{m}_t$ and $\hat{v}_t$ are the bias-corrected estimates of the first and second moments of the gradient. We used the learning rate scheduler \texttt{CosineAnnealingWarmRestarts} \cite{paszke2019pytorchimperativestylehighperformance}, which progressively adjusts the learning rate during training as

\begin{equation}
\alpha_t = \alpha_{\min}
+ \tfrac{1}{2}\bigl(\alpha_{\max} - \alpha_{\min}\bigr)
\left( 1 + \cos\!\left( \pi\frac{T_{\mathrm{cur}}}{T_i} \right) \right).
\end{equation}

where $\alpha_{\max}$ is the initial learning rate, $T_{\mathrm{cur}}$ is the number of epochs since the last restart, and $T_i$ is the period, i.e., the number of epochs between two warm restarts in stochastic gradient descent with warm restarts (SGDR)~\cite{loshchilov2017sgdrstochasticgradientdescent}. The initial value of $T_i$ was set to 20 epochs, and the period was doubled after each restart.

\subsection{Area Model}
The area model constitutes a comparatively simpler inference problem and therefore admits a substantially reduced network architecture. The critical scale for this task is the signal area corresponding to the TE polarization of the proton, which represents the smallest physically relevant signal amplitude encountered in practice. Achieving low relative error at this scale ensures quality calibration and robust performance across all higher-polarization regimes, where signal amplitudes are correspondingly larger.

We therefore use the area model as a controlled test case to examine how generalization performance evolves with model complexity in a relatively low-dimensional NMR problem. Based on this study, we find that a basic MLP architecture consisting of two hidden layers with 20 neurons each is sufficient to meet the accuracy and precision requirements of this task.

\subsection{Denoising Autoencoder}

\begin{figure}[h!]
    \centering
    \includegraphics[width=0.9\linewidth]{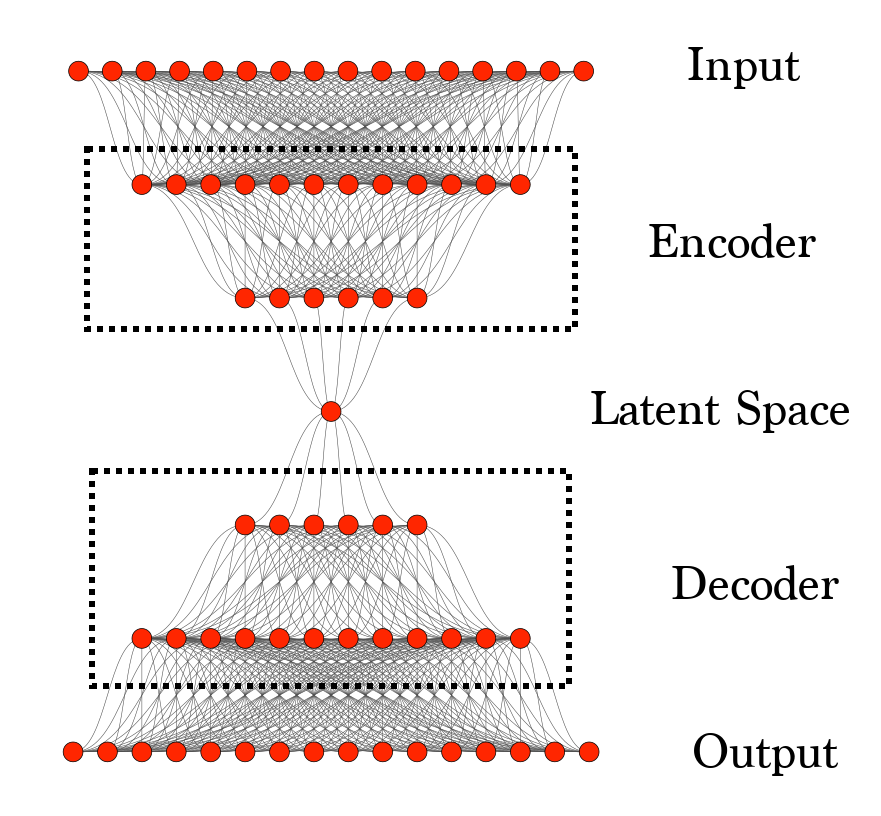}
    \caption{Example diagram of a denoising autoencoder architecture (not to scale). Diagram made with NN-SVG \cite{LeNail2019}.}
    \label{fig:autoencoder}
\end{figure}

In parallel, we developed a denoising autoencoder (DAE) \cite{michelucci2022introductionautoencoders,autoencoder-vincent} designed to filter noisy signals and reconstruct the underlying lineshape. The autoencoder consists of an encoder function \( f_\theta(\cdot) \) that maps an input signal \( x \in \mathbb{R}^{N} \) into a compressed latent representation \( z \in \mathbb{R}^{d} \), and a decoder \( g_\phi(\cdot) \) that reconstructs a clean estimate \( \hat{x} \):
\begin{equation}
    z = f_\theta(x), \qquad \hat{x} = g_\phi(z).
\end{equation}
A schematic diagram of a simplified denoising autoencoder (DAE) architecture is shown in Fig.~\ref{fig:autoencoder}. This diagram is intended solely as a conceptual representation; the full network used in this work comprises significantly higher-dimensional layers and cannot be depicted compactly without loss of clarity.

The input to the DAE consists of a frequency-domain NMR spectrum with a feature dimension of 500, corresponding to the number of frequency bins. The output layer is constructed with the same dimensionality to enable direct reconstruction of the denoised spectrum. The hidden layers implement a symmetric encoder--decoder structure, in which the feature dimensionality is progressively reduced from 500 down to a low-dimensional latent representation of size 4 and subsequently expanded back to 500. This compression and reconstruction are performed through a sequence of reductions and expansions by factors of two, resulting in a total of 12 hidden layers. Rectified linear unit (ReLU) activations are used between all intermediate layers to introduce nonlinearity, while a linear activation function is applied at the output layer to preserve the continuous-valued nature of the reconstructed signal.

To train the network in a denoising configuration, noise \( \varepsilon \sim \mathcal{D}_\varepsilon \) is added to the clean signal to form a corrupted input \( \tilde{x} = x + \varepsilon \). The DAE is optimized to minimize reconstruction error between \( \hat{x} = g_\phi(f_\theta(\tilde{x})) \) and the clean target \( x \) using a mean-squared error (MSE) loss:
\begin{equation}
    \mathcal{L}_{\text{DAE}}(\theta,\phi) = \frac{1}{N} \sum_{i=1}^{N} \left\lVert x_i - g_\phi \!\left( f_\theta(\tilde{x}_i) \right) \right\rVert_2^2 .
\end{equation}
To preserve physically meaningful spectral structure (such as smooth curvature and peak shape), we include a second-derivative regularization penalty:
\begin{equation}
    \mathcal{L}_{\text{smooth}} = \left\lVert \nabla^{2} g_\phi \!\left( f_\theta(\tilde{x}) \right) \right\rVert_2^2 .
\end{equation}
The total loss is therefore:
\begin{equation}
    \mathcal{L} = \mathcal{L}_{\text{DAE}} + \lambda \, \mathcal{L}_{\text{smooth}},
\end{equation}
where \( \lambda \) controls the trade-off between target fidelity and smoothness. The trained autoencoder then acts as a noise-removal mapping,
\begin{equation}
    \hat{x} = \mathcal{F}_{\theta,\phi}(x + \varepsilon).
\end{equation}


\subsection{Training}
\label{sec:training}
Training was performed using datasets generated through MC simulations. For each simulated event, a baseline parameterization is randomly sampled according to the formalism described in Sec.~\ref{sec:qmeter}, after which the NMR signal is constructed using the appropriate lineshape model. Various scales of white and structured noise (as described in Sec \ref{sec:errors}) are added independently to each of the 500 frequency bins to emulate electronic and RF noise. The resulting composite spectrum is stored as a simulated polarized-target NMR event, representing a realistic approximation of an experimentally recorded NMR spectrum obtained from the accumulation of multiple NMR sweeps.

Validation and testing were then performed on both experimentally representative noise conditions and more challenging simulated cases with increased noise complexity and structured baseline distortions, allowing the robustness and generalization of the models to be assessed.

For the area model, simulated signals were generated over the full polarization range of 100\%. Polarization values near the proton thermal-equilibrium (TE) region, up to 1\%, were intentionally oversampled using a uniform distribution, while values above 1\% were sampled according to an inverse-exponential distribution. This sampling strategy increases the statistical weight of low-polarization events, which would otherwise be underrepresented in a purely uniform dataset, and ensures that the model learns to resolve the smallest physically relevant signal areas with high fidelity.

By emphasizing this low-area regime during training, the model is forced to develop sensitivity to subtle variations in signal area that dominate the error budget near TE, while naturally retaining strong performance at higher polarization values where the signal-to-noise ratio and feature contrast are larger.

For the DAE, training data were generated over a polarization range of 
0--60\%, with the region near the TE polarization up to 1\% intentionally oversampled to emphasize the most noise-sensitive regime. Because the DAE is tasked with learning an implicit representation of the underlying physical lineshape through reconstruction, stochastic noise was injected into the input spectra while the reconstruction targets were kept clean. This approach forces the encoder to learn a compact latent representation that captures invariant spectral structure while discarding noise-dependent fluctuations.

For validation and testing, the clean target spectra were kept noiseless, while corrupted inputs were generated with independent noise realizations when denoising performance was evaluated. This choice ensures that loss evaluation probes the fidelity with which the latent space encodes the intrinsic NMR lineshape, rather than rewarding overfitting to a specific noise realization. By separating the clean targets from independently corrupted inputs, the training procedure preserves the embedded physical constraints of the reconstruction task and yields a latent representation that generalizes robustly across noise conditions encountered during inference.

\begin{figure}
    \centering
    \includegraphics[width=\linewidth]{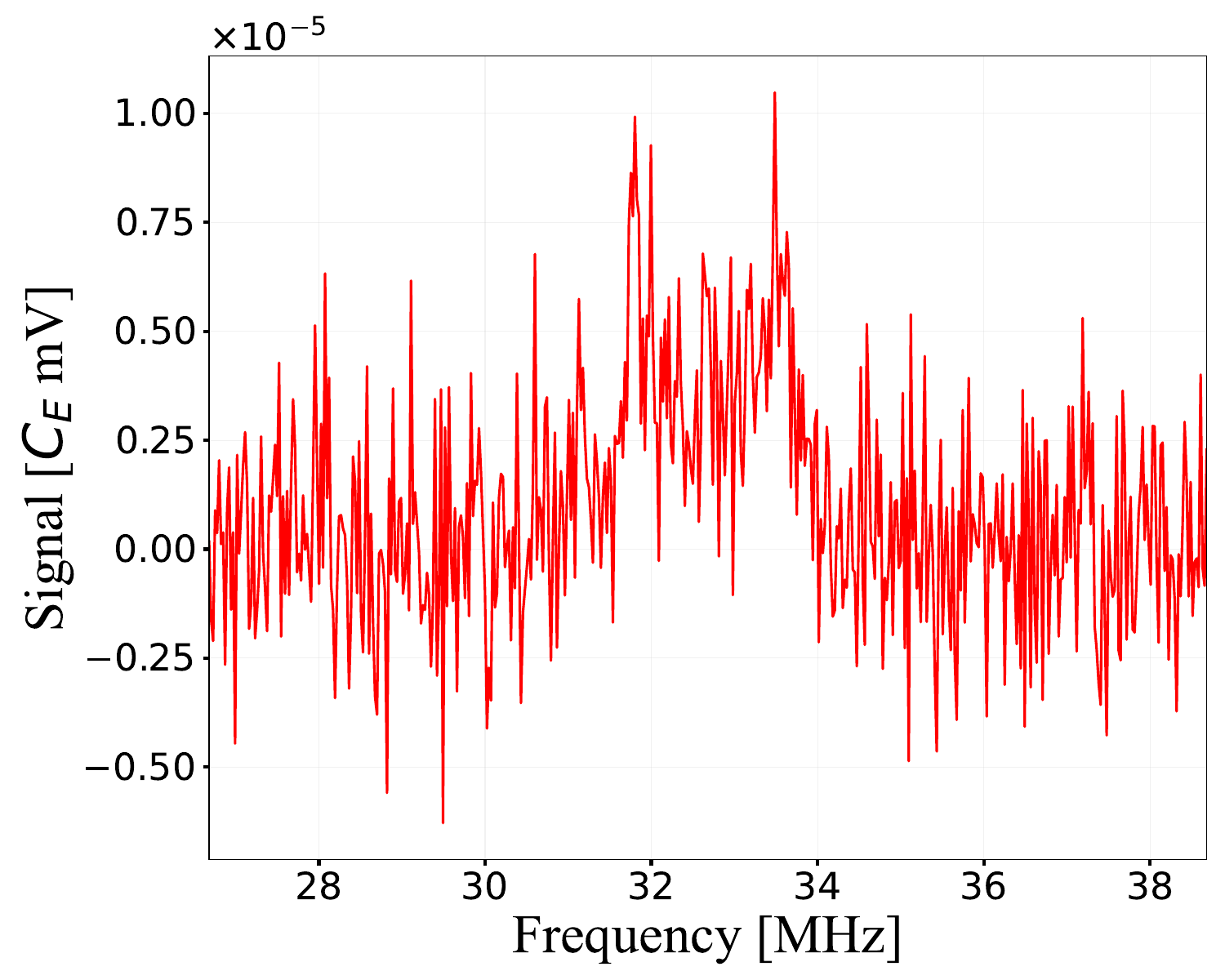}
    \caption{Simulation of a TE signal with the higher injected noise scale. For this specific TE example, the SNR is 1.11, corresponding to approximately double the nominal noise-amplitude scale shown in Fig.~\ref{fig:D-Baseline-TE}.}
    \label{fig:placeholder}
\end{figure}

Training, validation, and testing datasets were split according to an 80/10/10 ratio. All models were initially trained for 1000 epochs with a batch size of 256. If the validation loss continued to decrease, the model weights were reloaded and training was resumed until overfitting began to appear. Model was constructed and trained using the \texttt{PyTorch} API~\cite{paszke2019pytorchimperativestylehighperformance} with the \texttt{AdamW} optimizer~\cite{loshchilov2019decoupledweightdecayregularization}. In general, each model should ideally be trained on a dataset containing at least an order of magnitude more independent training samples than the number of free trainable parameters, with the precise requirement depending on the effective model complexity; this rule of thumb is not strictly satisfied for every benchmark presented here. Unless otherwise stated, each initial benchmark model was trained using one million simulated events, corresponding to deuteron NMR signals superimposed on realistic Q-meter baselines and spanning a broad range of polarization values and noise conditions.

Hyperparameters were optimized using \texttt{Optuna}~\cite{akiba2019optunanextgenerationhyperparameteroptimization}, a Bayesian optimization framework that performs iterative trials to identify optimal hyperparameter configurations within predefined constraints. We ran 200 trials per model to determine near-optimal configurations, except for the low-polarization model, for which we used the same optimized hyperparameters as the high-polarization model. These tuned hyperparameters are shown in Tables~\ref{tab:ff-parameters} and~\ref{tab:polarization_parameters}. During training, the nominal injected noise scale was chosen such that a representative TE deuteron event has SNR $=2.6$, as shown in the bottom panel of Fig.~\ref{fig:D-Baseline-TE}. A doubled noise-amplitude scale, corresponding to SNR $=1.3$ for that same representative event, was used for more extreme validation (see Fig.~\ref{fig:placeholder}). The noise was sampled from a normal distribution $\mathcal{N}(0,\sigma)$, where the mean is zero and $\sigma$ is the standard deviation. From experimental signals, this standard deviation was found to be on the order of $10^{-6}$~mV. Discussions of ``high noise'' regions in this paper refer to noise sampled from a normal distribution with twice this standard deviation; the resulting event-level SNR depends on the signal amplitude through Eq.~\ref{eqn:SNR}.

Model training is ideally performed either in a single uninterrupted session or through the use of checkpointing, such that the optimized model state and optimizer parameters are preserved and training can resume seamlessly from the point of interruption. This approach is particularly important for models that require extended training times or are trained on large datasets.

\begin{figure}[h!]
    \centering
    \includegraphics[width=0.5\textwidth]{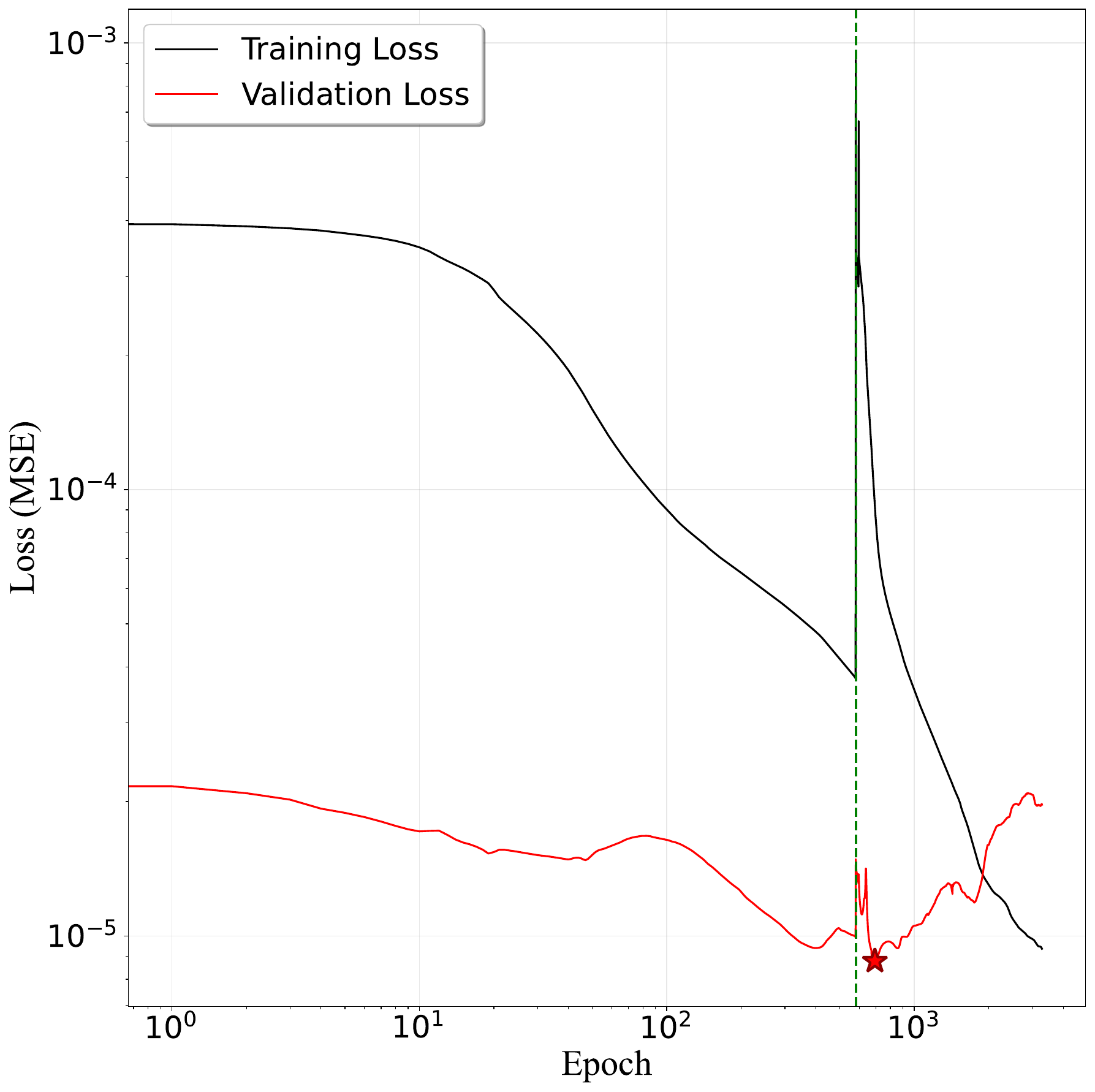}
    \caption{Loss/Validation Loss vs. Epoch for area model (log-log scaling). Early stopping was applied when the model started to overfit (i.e., after 500 epochs without any improvement). The red star indicates the lowest validation loss achieved; the model weights and biases at this point were used as the final model.}
    \label{fig:Area-Loss}
\end{figure}

Figure~\ref{fig:Area-Loss} shows a representative training history for the area model. The spike marked by the green vertical dotted line corresponds to a point at which the model weights were reloaded without restoring the associated optimizer state; as a consequence, the optimizer was effectively reinitialized, producing a transient increase in the loss. The red star denotes the epoch at which the minimum validation loss was achieved, corresponding to the optimal set of model weights and biases, which were subsequently saved as the final trained model. For all CNN models, the simulated events were generated for the training procedure with the dataset partitioned according to an 80/10/10 split into training, validation, and testing subsets.

\section{Results}
\label{sec:results}

In this section, we present the evaluation of the four specialized model classes introduced above. Their performance is assessed using independently generated test data that systematically vary in polarization, phase mixing, lineshape, baseline drift, and SNR. This allows us to characterize the strengths and limitations of each model architecture---MLP, CNN, DAE, and the hybrid variants---in terms of accuracy, precision, stability under distribution shifts, and sensitivity to both statistical and systematic variations in the underlying NMR signals. Studies using real experimental noise were also performed, but they proved too simple compared with the more flexible noise-injection model used for the benchmarks. The simulated noise includes the components described in Sec.~\ref{sec:errors}.

Both the polarization and area models were evaluated using simulated NMR spectra generated with the simulated baseline and signal lineshape formalism described in Secs.~\ref{sec:qmeter} and~\ref{sec:spin-1}, respectively. For each simulated target event we passed the resulting spectrum to the model to obtain inferred values of polarization, area, or NMR signal in the case of the DAE. This procedure was repeated many times (typically 1000 trials) across the full polarization range or the specialized range of a given model to quantify both the mean and spread of the distribution of the residuals.

\subsection{Standard Fits}

An error analysis of the standard non-AI lineshape-fitting procedure, based on least-squares \(\chi^2\) minimization, was performed using the same simulated datasets employed to benchmark the AI models. The procedure consists first of baseline subtraction, in which a third-order polynomial is fit by \(\chi^2\) minimization to the spectral wings and then subtracted to isolate the target signal. A Dulya-type fit \cite{Dulya1997} is then applied to extract the polarization.

The quality of the fit is strongly polarization dependent, with higher polarization values yielding substantially more stable and accurate extractions. To enable a direct comparison with the AI-based models, the analysis was performed over the same polarization ranges with the same test data. For polarization values between \(10\%\) and \(60\%\), the average relative extraction error was approximately \(3.5\%\) over 1000 trials.

This fitting procedure was fully automated. As a result, the baseline subtraction and signal alignment were not optimized on a case-by-case basis, as would normally be required in a careful offline analysis. In particular, no manual adjustments were made to ensure that offset signals were optimally positioned within the fit function or that the spectral wings were selected to minimize the baseline-subtraction contribution to the \(\chi^2\). To estimate the improvement achievable with expert intervention, a subset of 20 trials was analyzed with a human-in-the-loop procedure, in which special care was taken to obtain the best fit in each case. This reduced the relative extraction error in the same polarization region to slightly above \(1\%\).

Below \(10\%\) polarization, the extraction error increases rapidly as the signal amplitude decreases, especially under reduced signal-to-noise conditions. In the TE regime, the relative error can quickly approach \(100\%\) when the noise level is high or when the fit quality is poor. For nominal noise levels \((\mathrm{SNR}\approx 2)\) the lowest relative errors obtained were on the order of \(10\%\). However, this level of performance was achieved only for approximately \(10\%\) of the 1000 generated trials, and only when special care was taken to obtain the best possible fit.  On average, the fit error is simply too large to make a meaningful comparison.

A comparable evaluation was also performed for the area-based extraction method using proton-like NMR signals. In this case, the procedure follows the same initial baseline-subtraction approach: a third-order polynomial is fit by \(\chi^2\) minimization to the spectral wings and subsequently subtracted from the spectrum. The isolated signal is then numerically integrated using a Riemann-sum method \cite{KELLER2013133}, and the polarization is extracted from the resulting signal area.

At the TE polarization scale, corresponding to \(P \simeq 0.3\%\), the average relative extraction error was approximately \(2.5\%\) for the fully automated analysis. When the same procedure was applied to a subset of 20 trials with human-in-the-loop optimization, in which special care was taken to improve the baseline fit and integration window on a case-by-case basis, the relative extraction error was reduced to approximately \(0.7\%\).

\subsection{Polarization Models}
\label{sec:polarization-models}

Across the high-polarization regime, \(P = 2\%\)--\(60\%\), the basic feedforward MLP achieved a mean residual of \(0.107\%\) and a residual spread of \(0.150\%\). This corresponds to an effective polarization-extraction uncertainty of approximately \(0.3\%\) relative at the upper end of the polarization range. The fully optimized CNN architecture achieved an additional reduction of roughly an order of magnitude in the relative extraction error, rendering this contribution negligible for practical polarization measurements. The noise range used in this study corresponds to signal-to-noise ratios of approximately \(\mathrm{SNR}=5\)--\(10\).

The CNN-based model developed for the low-polarization regime TE--\(2\%\) significantly outperforms the corresponding MLP-based architecture. When evaluated on noiseless test data, the CNN yields a mean residual of \(3\times10^{-5}\%\) with a residual standard deviation of \(2.3\times10^{-4}\%\), demonstrating excellent intrinsic accuracy and precision for such a small signal scale. This corresponds to roughly 0.4\% relative error.

Under noisy conditions with signal-to-noise ratios in the range \(\mathrm{SNR} \approx 5\)--\(10\), the model maintains excellent performance, achieving a mean residual of \(7.00 \times 10^{-4}\%\) and a residual spread of \(2.03 \times 10^{-3}\%\), corresponding to roughly 4\% relative error across all 1000 trials. Even in more challenging conditions (\(\mathrm{SNR} \approx 1\)--\(5\)), the CNN-based model remains robust, yielding a mean residual of \(-2.31 \times 10^{-3}\%\) and a precision of \(5.08 \times 10^{-3}\%\).

For the most direct comparison, we used a dataset with a fixed SNR of 2. Table~\ref{tab:accuracy_precision} summarizes the overall performance of the least-squares fit (top 1\% of 1000 trials) compared with the MLP and CNN models (all 1000 trials).
\begin{table}[htbp]
    \centering
    \begin{tabular}{lcc}
        \hline
        \textbf{Model} & \textbf{$\Delta P/P$} \\
        \hline
        CNN & 3.8\% \\
        MLP & 8.2\%  \\
        $\chi^2$-fit & 16\% \\
        \hline
    \end{tabular}
    \caption{Summary of best results (relative error) for fully automated extraction using 1000 trials of TE-scale deuteron signal at SNR of 2 using the CNN, the basic MLP, and a standard least-squares fit (top 1\%).}
    \label{tab:accuracy_precision}
\end{table}

\begin{figure}[h]
    \centering
    \includegraphics[width=0.5\textwidth]{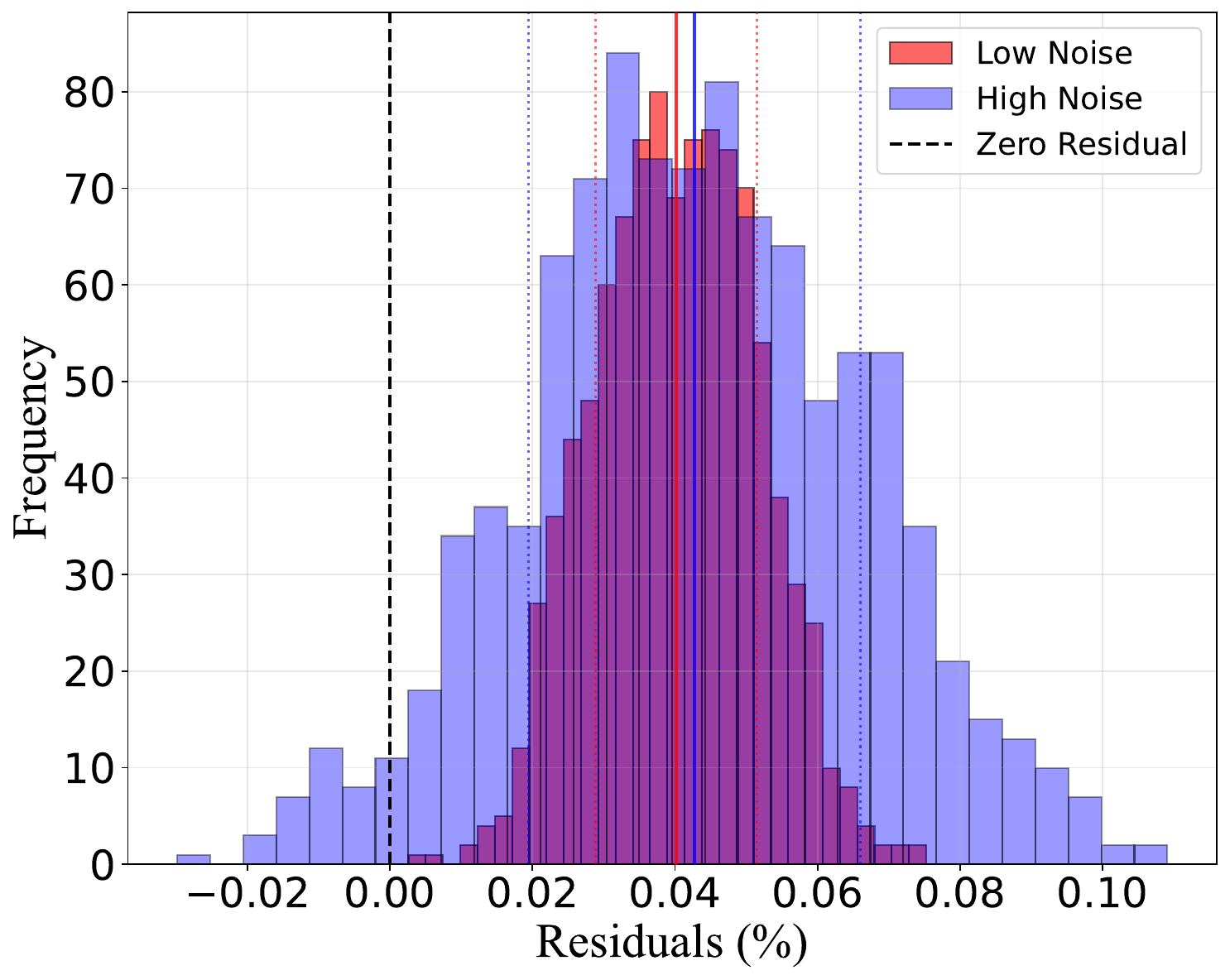}
    \caption{Distribution of residuals for polarization of $P=5\%$ from 1000 inference tests with the high--polarization model. \textbf{Red}: The noise applied gives an SNR around 10. \textbf{Blue}: The same noise (doubled), giving an SNR of 5.}
    \label{fig:SideBySideComparison_polarization}
\end{figure}

Because the training data are uniformly distributed in polarization, model accuracy and precision naturally vary across different test regions. This behavior primarily reflects changes in the NMR lineshape as a function of polarization: depending on the local structure of the lineshape, a model may exhibit a trade-off between bias (accuracy) and variance (precision) in specific polarization intervals, while maintaining roughly constant overall performance (see Table~\ref{tab:residuals}). An example is shown in Fig.~\ref{fig:SideBySideComparison_polarization}, which displays residuals from 1000 inference trials at a fixed true polarization of 5\% under varying noise and baseline conditions. Two representative noise regimes are shown: a lower-noise case (SNR = 10) and a higher-noise case (SNR = 5).

\begin{table}[h]
\centering
\small
\begin{tabular}{llcc|cc}
\toprule
\textbf{Range} & \textbf{Noise} 
& \multicolumn{2}{c|}{\textbf{$\chi^2$ Fit Res.}} 
& \multicolumn{2}{c}{\textbf{DNN Res.}} \\
\cmidrule(lr){3-4}\cmidrule(lr){5-6}
 &  & \textbf{$\mu$ (\%)} & \textbf{$\sigma$ (\%)} 
    & \textbf{$\mu$ (\%)} & \textbf{$\sigma$ (\%)} \\
\midrule

10--15\% & Low 
& 0.327 & 0.901 
& 0.010 & 0.061 \\

10--15\% & High
& 0.362 & 1.832 
& 0.015 & 0.063 \\

20--25\% & Low  
& 0.260 & 0.646 
& 0.006 & 0.055 \\

20--25\% & High 
& 0.266 & 1.297 
& 0.009 & 0.056 \\

30--35\% & Low 
& 0.248 & 0.572 
& 0.003 & 0.051 \\

30--35\% & High
& 0.237 & 1.146 
& 0.004 & 0.057 \\

40--45\% & Low 
& 0.259 & 0.533 
& 0.002 & 0.052 \\

40--45\% & High 
& 0.244 & 1.063 
& 0.003 & 0.060 \\

50--55\% & Low
& 0.259 & 0.501 
& 0.001 & 0.065 \\

50--55\% & High
& 0.241 & 0.998 
& 0.001 & 0.065 \\
\bottomrule
\end{tabular}
\caption{Comparison of residual statistics for our best standard least-squares $\chi^2$ fit and our best DNN model extraction on the same data across different polarization scales and noise levels for the deuteron signal type. All residual means and standard deviations are absolute residuals reported in units of polarization percent. Results are shown for testing the 2--60\% model for low noise (13--476 signal-to-noise) and high noise (6--180 signal-to-noise) ranges.}
\label{tab:residuals}
\end{table}

\subsection{Area Model}
We first examined the area-extraction problem under idealized conditions in order to establish a benchmark for achievable performance. In the absence of frequency-dependent distortions---specifically when the proton NMR peak position is fixed and the Q-meter baseline exhibits neither tune shifts nor discontinuous jumps---the relationship between integrated signal area and polarization is approximately linear over the full dynamic range. Under these conditions, a simple linear ridge regression model is sufficient to achieve a relative error of approximately $0.12\%$ in the TE region for the proton.

Here, the ridge model refers to a linear regression augmented with an $\ell_2$ regularization term on the model coefficients. This regularization suppresses sensitivity to noise and mitigates overfitting when training data are limited, while preserving the underlying linear mapping between signal area and polarization. Owing to the near-linearity of the area--polarization relationship, this approach remains effective even when trained on a relatively small dataset of $\sim5\times10^3$ events under nominal noise conditions.

When extending the training to more realistic experimental scenarios---including variations in the NMR peak position, baseline tune shifts, and small discontinuities in the Q-curve---the effective mapping between signal area and polarization becomes nonlinear. In this regime, a purely linear model is no longer sufficient, and sub-percent relative error at the TE scale cannot be maintained. Introducing modest nonlinearity by augmenting the model with two hidden layers of 20 neurons each restores performance, allowing the model to again achieve relative errors below $0.7\%$ in the TE region, provided that the training dataset is increased to at least $\sim5\times10^4$ events.

For completeness, we also evaluated a CNN architecture---identical in form to that used for the polarization models---on the area-extraction task. While the CNN is capable of achieving comparable relative error, doing so requires substantially larger training datasets, on the order of $10^6$ events. This reflects the greater model capacity and weaker inductive bias of the CNN for a problem that is fundamentally low dimensional and dominated by global amplitude information rather than localized spectral features.

\begin{figure}[h]
    \centering
    \includegraphics[width=0.5\textwidth]{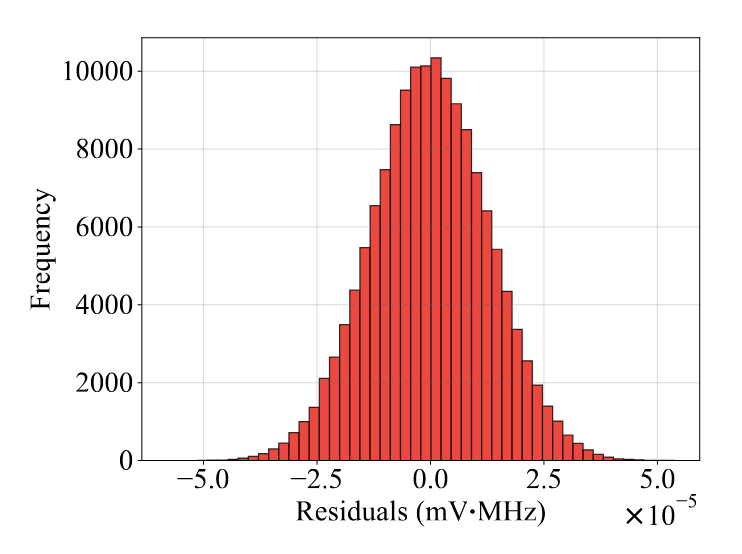}
    \caption{Distribution of residuals in the area model for the polarization range from TE to 60\%. The SNR range for this test is 6--77.}
    \label{fig:Area-Residuals}
\end{figure}

Figure~\ref{fig:Area-Residuals} shows the distribution of residuals between the predicted and true signal area for the MLP-based area model across the full tested polarization range for a proton-like signal. For the independent inference trials used to generate the histogram, the mean residual is $2.9855\times10^{-8}\,\mathrm{mV}\!\cdot\!\mathrm{MHz}$, with a residual width of $1.288\times10^{-5}\,\mathrm{mV}\!\cdot\!\mathrm{MHz}$. 

\begin{figure}[h]
    \centering
    \includegraphics[width=0.5\textwidth]{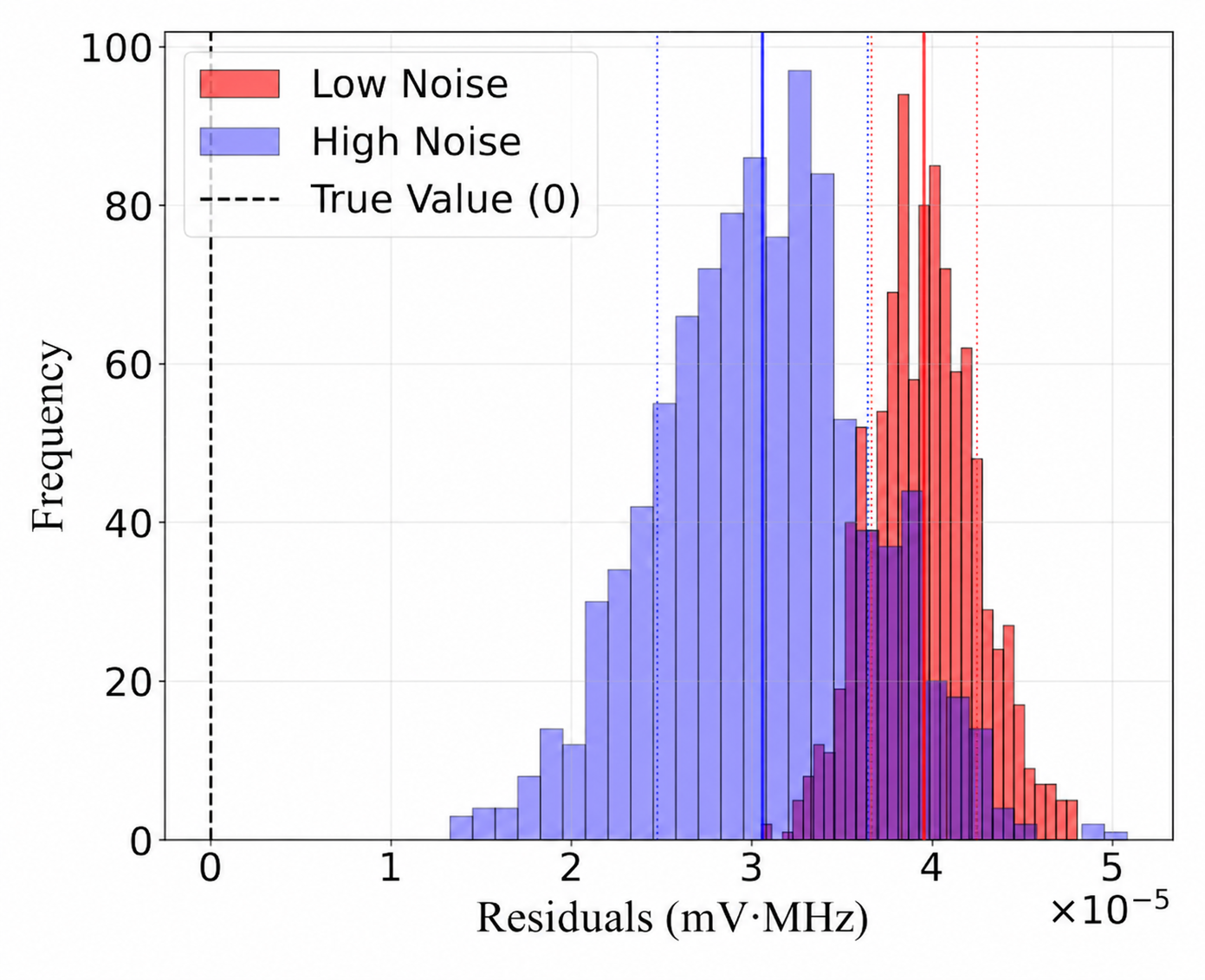}
    \caption{Distribution of residuals from the area model for a polarization of $P=5\%$ over 1000 trials of inference for the polarization range from TE to 60\%. For this test, the SNR range for the low-noise case (red) was $\sim$30 and for the high-noise case (blue) was $\sim$15. The mean and standard deviation are shown in Table~\ref{tab:residual_stats}.}
    \label{fig:SideBySideComparison_area}
\end{figure}

The area model successfully generalized across realistic lineshapes by inferring only the total area under the curve.  
Fig.~\ref{fig:SideBySideComparison_area} shows the results for a deuteron-type Pake doublet lineshape area test.  The figure presents overlapping histograms of the residual distributions obtained under low-noise (SNR $=30$) and high-noise (SNR $=15$) conditions. These distributions were generated by repeatedly injecting independent noise realizations into a fixed underlying signal over 1000 trials, enabling a systematic characterization of noise-induced error. The corresponding mean and standard deviation values are summarized in Table~\ref{tab:residual_stats}.  Assuming zero error from the calibration constant in the mapping from area to polarization the area residuals can be converted to polarization residuals so a relative error in polarization can be calculated (also provided in Table \ref{tab:residual_stats}). Even in the high-noise regime, the resulting residuals correspond to a relative error of approximately $0.6\%$, demonstrating the robustness of the trained model.

\begin{table}[h]
\centering
\caption{Residual statistics for polarization ($P$) and area ($A$) for an example at $P = 5\%$.}
\begin{tabular}{lccc}
\hline
 & Noise & $\mu$ & $\sigma$ \\
\hline
$P (\%)$ & Low  & $0.010$ & $0.011$ \\
    & High & $0.013$ & $0.023$ \\
$A~(\mathrm{mV}\!\cdot\!\mathrm{MHz})$ & Low  & $4\times10^{-5}$ & $3\times10^{-6}$ \\
    & High & $3\times10^{-5}$ & $6\times10^{-6}$ \\
\hline
\end{tabular}
\label{tab:residual_stats}
\end{table}

\subsection{Denoising Autoencoder Model}

\begin{figure}[h!]
    \centering
    \includegraphics[width=\linewidth]{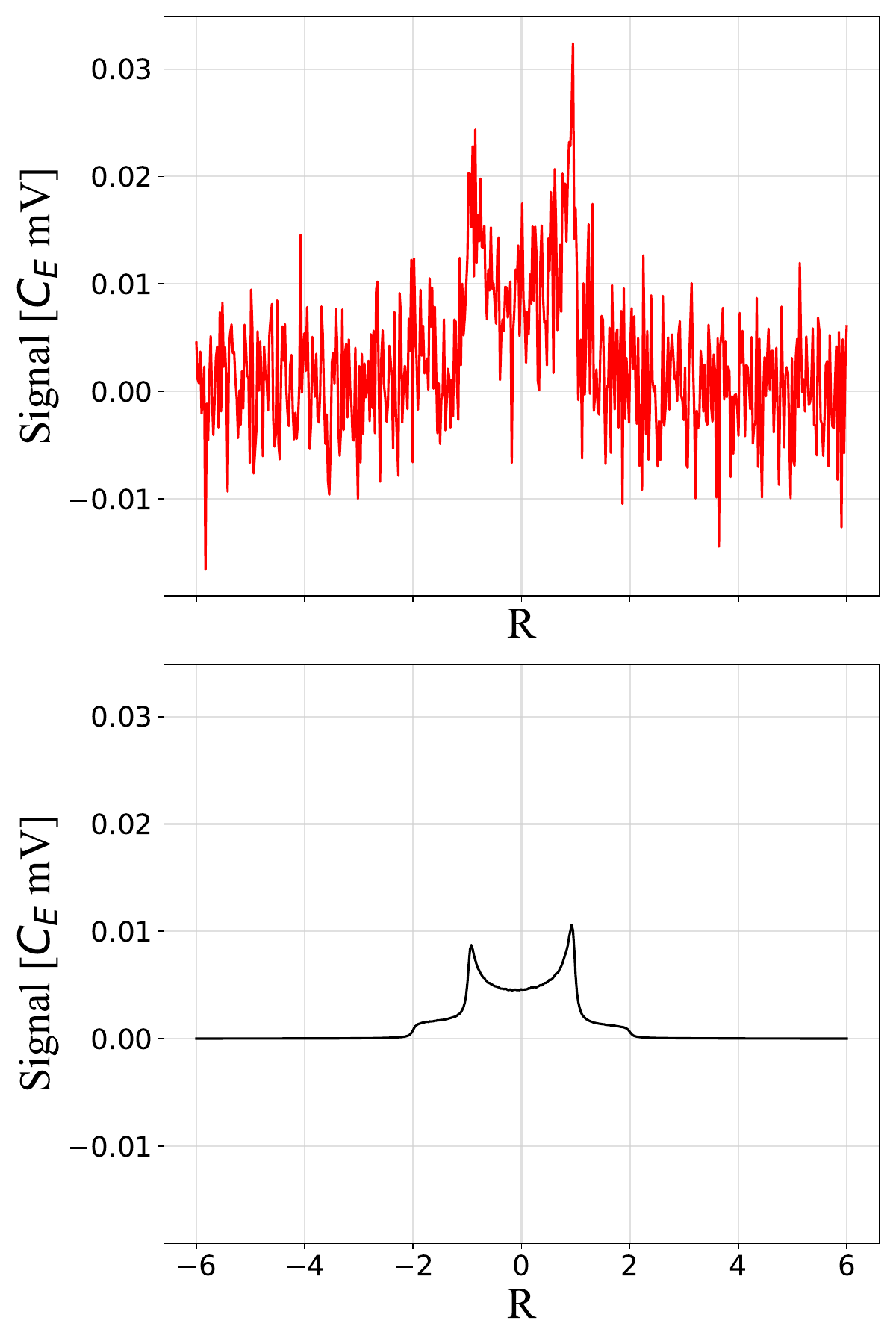}
    \caption{Example use of the DAE on a simulated lineshape event. \textbf{Top}: Input signal with a large amount of noise. \textbf{Bottom}: Reconstructed signal. SNR for this example is $0.8045$.}
    \label{fig:denoiser}
\end{figure}

\begin{figure}
    \centering
    \includegraphics[width=\linewidth]{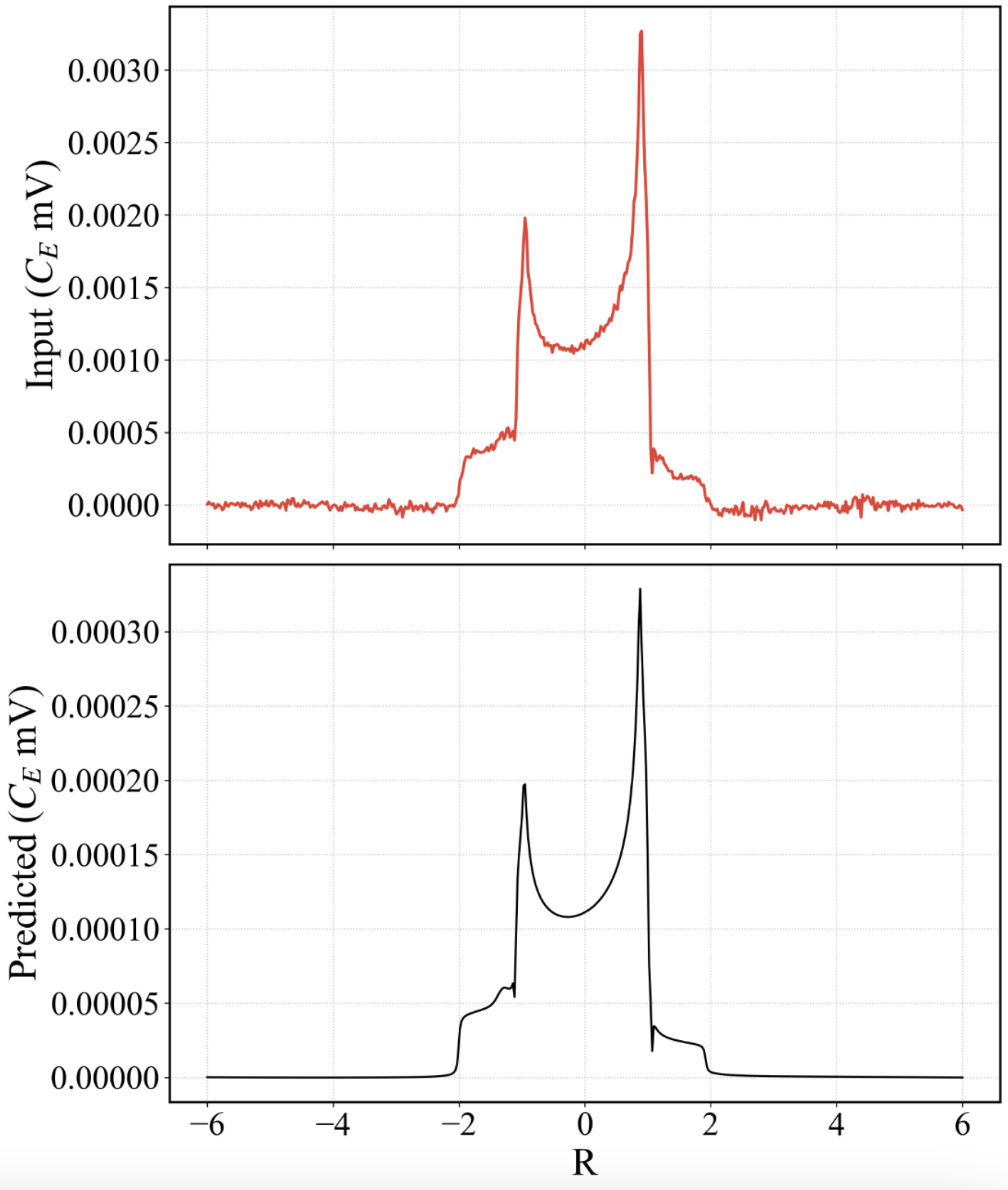}
    \caption{Example of the DAE applied to a real experimental signal under typical laboratory noise conditions for a single-sweep NMR measurement. The signal was acquired from irradiated d-butanol with an estimated polarization of $44\pm2\%$.}
    \label{fig:dae-real}
\end{figure}

\begin{figure}
    \centering
    \includegraphics[width=\linewidth]{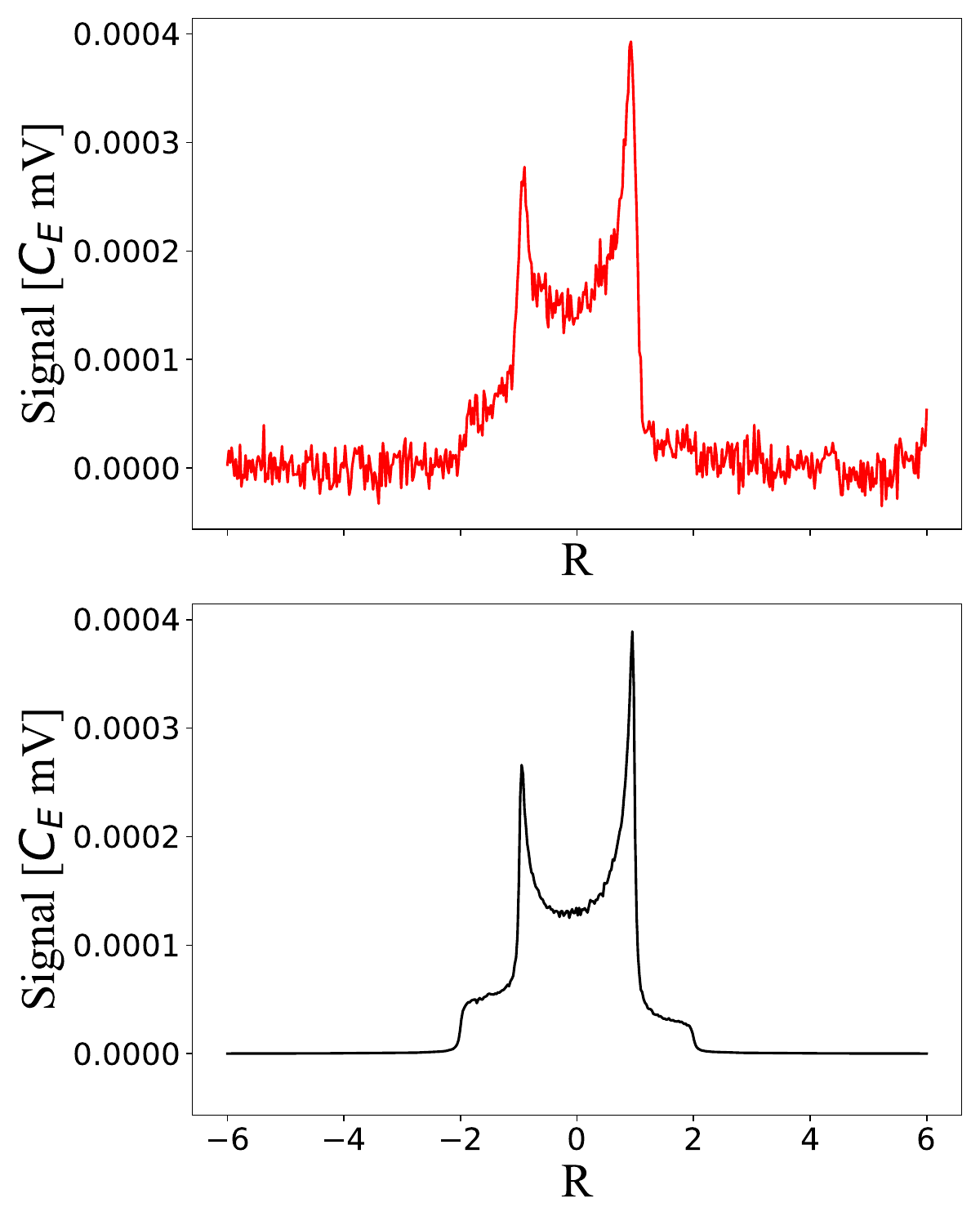}
    \caption{Another example of the DAE applied to real experimental data with the coil outside the target material, producing a significantly worse SNR. The signal shown here was acquired from an irradiated d-butanol sample with an estimated polarization of approximately 45\%.}
    \label{fig:dae-real_2}
\end{figure}

The denoising autoencoder (DAE) model is designed to suppress noise while preserving the underlying NMR signal structure. In this architecture, the input to the model consists of a noisy frequency-domain spectrum comprising 500 frequency bins, and the output is a spectrum of identical dimensionality in which the noise component is substantially reduced.

Figures~\ref{fig:denoiser} and~\ref{fig:dae-real} show representative examples of the DAE's performance on simulated and experimental spectra, respectively. Figure~\ref{fig:denoiser} illustrates denoising of a simulated NMR lineshape at a signal-to-noise ratio of approximately unity, $\mathrm{SNR}\approx 0.8$, corresponding to a polarization level of $5\%$. Because the clean simulated target spectrum is known exactly, this case provides a direct test of the reconstruction error. The residuals between the clean target spectrum and the DAE-reconstructed spectrum are very small, with a mean residual of $\mu = 5\times10^{-8}$ and a residual width of $\sigma = 1.5\times10^{-6}$ in the plotted signal units.  The DAE output can then be used as input to a trained polarization model to obtain $4.98\pm0.4\%$. This demonstrates that, under matched simulation conditions, the DAE can strongly suppress noise while preserving the dominant Pake-doublet structure.

Figure~\ref{fig:dae-real} shows the same DAE procedure applied to real experimental data recorded under typical laboratory noise conditions. For this spectrum, a standard Dulya-type fit yielded a polarization value of $44.2\%\pm2.3\%$, which provides the conventional-analysis reference for comparison. The denoised output from the DAE was then passed to a highly tuned d-butanol DNN polarization model, resulting in an inferred polarization of $43.8\%\pm0.03\%$. The two central values are consistent within the uncertainty of the conventional fit. The quoted uncertainties should be interpreted as extraction uncertainties associated with the respective analysis procedures and assumptions, not as the full experimental polarization uncertainty. Here, ``highly tuned'' means that the DNN was trained on simulated spectra whose functional form and material-specific artifacts, including the additional structure associated with the O-D bond in d-butanol, were carefully matched to the experimental lineshape.

In addition, Fig.~\ref{fig:dae-real_2} shows a deliberately challenging application of the DAE to experimental data acquired with the NMR coil displaced approximately 10~mm from the target cup to allow mechanical rotation. This geometry reduces the effective coil--sample coupling and filling factor, thereby lowering the signal amplitude and degrading the apparent signal-to-noise ratio relative to the configurations represented in the training set. The reconstruction was therefore evaluated as an out-of-distribution test of the model's ability to generalize beyond the nominal training conditions. Despite this mismatch, the DAE suppresses much of the high-frequency noise while preserving the dominant spectral structure and peak positions of the d-butanol signal. Only modest distortions of the reconstructed lineshape are visible, illustrating the potential utility of DAE-type architectures for noise reduction under degraded coupling conditions. 

As a qualitative robustness check, the denoised spectrum was also passed through the downstream polarization-extraction model. For this example, the DAE/DNN analysis gives an inferred polarization of $42.3\pm0.2\%$, where the quoted uncertainty reflects the extraction uncertainty under the model assumptions rather than the full experimental polarization uncertainty. A conventional least-squares Dulya-type fit applied to the same spectrum gives $45.4\pm4.5\%$. The two estimates are therefore consistent within the substantially larger least-squares uncertainty, but the comparison should not be interpreted as a full validation of the DAE/DNN method in this displaced-coil geometry. Quantitative use in this regime would require either dedicated validation data or additional training samples generated with comparable coil--target coupling, filling factor, and noise conditions.

\subsection{Comparison}

For polarizations above a few percent, we observe that the extraction error can be effectively eliminated as a leading contribution, with AI-based extraction reducing the extraction-related contribution to the overall polarization error budget by 1–2\% in some cases. In the low-polarization regime, especially in the TE range for Pake-doublet signals, the DNN and CNN still contain residual extraction errors, but these errors are substantially smaller than those from standard extraction methods. This direct, one-to-one comparison with standard non-AI approaches demonstrates the performance gains afforded by the proposed AI-based techniques. Additionally, as noted above, the benchmarks use limited training datasets to reduce training time. For real-world deployment, larger and more fully tuned training ensembles should further improve performance and lower the overall extraction error.

An additional source of uncertainty, not quantified explicitly in this study, is the effect of covariate shift in the training data. This refers to the methodological error introduced when the simulated training distribution differs from the real experimental data to which the model is ultimately applied. Minimizing this contribution requires careful tuning and validation of the Monte Carlo training samples so that the simulated lineshapes reproduce, as accurately as possible, both the known theoretical response and the experimentally constrained parameterization under the relevant operating conditions.

For the controlled examples presented here, this contribution is expected to be negligible because the test samples are generated from the same underlying simulation model used for training. In a real experimental implementation, however, substantial effort may be required to minimize this component of the total extraction error. This includes tuning the simulated lineshape model, validating the noise and baseline structure, accounting for possible instrumental distortions, and ensuring that the training ensemble spans the range of experimental conditions encountered in the data.

Although these models achieve substantial reductions in extraction-related uncertainty, it is important to acknowledge instrumental limitations that the AI-based methods considered here cannot directly circumvent. In particular, reducing the extraction error to levels well below the Q-meter's intrinsic relative accuracy of approximately 
1\% does not translate into a proportional improvement in the overall measurement precision. While such reductions effectively render the fitting uncertainty negligible within the total error budget, a Q-meter-based NMR system cannot exceed this fundamental performance limit without corresponding advances in hardware design and instrumentation.

In addition, the achievable lower bound on the overall polarization uncertainty is constrained by several experimental factors. These include the coupling between the NMR coil and the target material, the spatial distribution of the beam over the sample, the associated nonuniform radiation damage, the distribution of microwave power throughout the target volume, and distortions in the magnetic-field homogeneity across the sample. The beam profile is especially important because it helps define the effective active target sampled by the experiment; if the beam distribution evolves with time, it can generate corresponding spatial variations in radiation damage, paramagnetic-center density, and polarization. This makes the relation between the NMR-measured average polarization and the polarization relevant to the scattering process inherently more complex. When TE calibration is used, uncertainty in the target temperature is also a critical contribution, since the TE polarization depends directly on temperature. Most notably, the inherently limited number of TE measurements that can be performed during an experiment, particularly for proton targets when relying on the area-based method, places a stringent bound on the extent to which offline analysis can further reduce the total uncertainty.

From the perspective of online monitoring and real-time feedback, inference speed constitutes an additional point of comparison between traditional and AI-based analysis methods. Conventional approaches---consisting of background subtraction, polynomial fitting and subtraction of the spectral wings, followed by numerical integration of the remaining signal---can typically be executed on time scales of a few hundred milliseconds per spectrum. This performance is generally adequate when many NMR sweeps are accumulated and averaged over time scales of seconds or more. However, such latencies are suboptimal for applications requiring rapid signal processing, adaptive response, or real-time system control.  Further, as demonstrated in our own standard least-squares fits, automating for speed without human quality checks on the fit substantially increases the expected fit error.

By contrast, all AI-based models considered here produce inference outputs on millisecond time scales on CPU systems, with further speed gains achievable through model optimization and hardware acceleration. This capability enables near-real-time analysis of individual or minimally averaged spectra. Moreover, for more sophisticated analyses---such as Dulya-type fits used to extract vector and tensor polarizations---there has traditionally been no comparably fast implementation on this time scale, since these procedures are computationally intensive and often require manual intervention to ensure stable convergence and reliable fit quality. Although Jefferson Lab has recently demonstrated online Dulya-type fitting on the scale of a few seconds, AI-based inference still offers a clear advantage in speed, scalability, and operational autonomy.

\section{Tensor Enhanced Measurements}
\label{sec:tensor}
Beyond vector polarization extraction, these same tools are expected to facilitate real-time tensor-polarization extraction once tensor-enhanced training labels and RF-manipulated lineshapes are incorporated. In such a model, the tensor polarization could be inferred directly from the trained lineshape response rather than being obtained only by propagating the vector-polarization uncertainty through the Boltzmann relation.
\begin{figure}[h!]
    \centering
    \includegraphics[width=\linewidth]{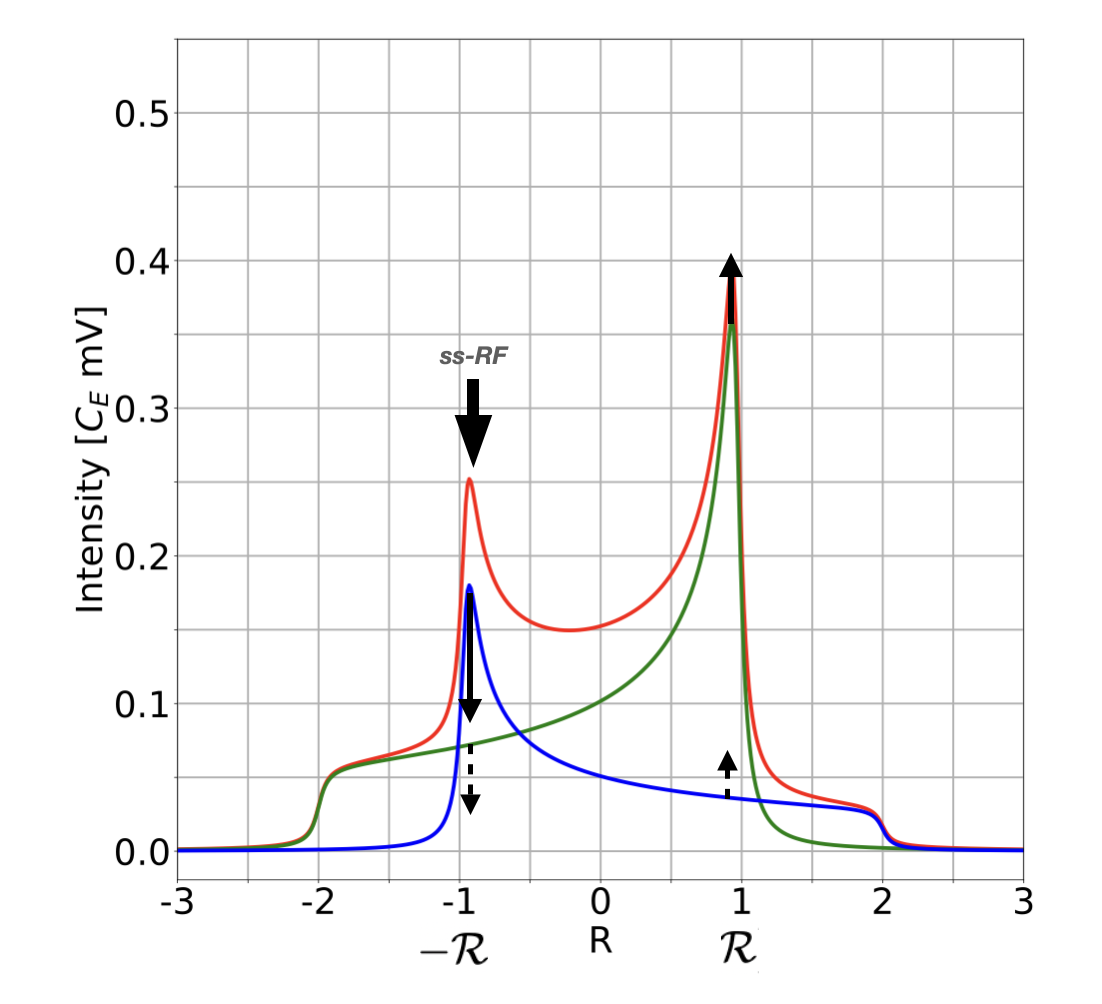}
    \caption{Application of ss-RF at the resonance position $-\mathcal{R}$ of the deuterium lineshape to reduce vector polarization \cite{CLEMENT2023168177}.}
    \label{fig:ss-RF}
\end{figure}

Incorporating ss-RF signal manipulations \cite{CLEMENT2023168177} directly into the simulation and training pipeline to obtain reliable extraction of the enhanced tensor polarization, as shown in Fig.~\ref{fig:ss-RF}, is the next natural step. To support this, we will investigate the use of differential binning to train the network to predict individual absorption components, enabling simultaneous extraction of both vector and tensor polarization from an RF-manipulated signal at millisecond time scales under various noise conditions. We also plan to explore physics-informed architectural constraints~\cite{2019JCoPh.378..686R}, including domain-aware regularization~\cite{stolte2023dominodomainawarelossregularization} and symmetry priors~\cite{bronstein2021geometricdeeplearninggrids}. When combined with simulated ss-RF perturbations over various frequency segments, we anticipate achieving a model capable of extracting high-fidelity polarization measurements under manipulations where conventional analysis methods fail.  This new analysis will be presented in future work.

\section{Conclusion}
\label{sec:conclusion}

In this work, we have demonstrated that deep neural network--based approaches provide a robust and practical extension to conventional CW-NMR polarimetry for dynamically polarized targets. By training on physically realistic simulations that incorporate baseline distortions, tune shifts, and multiple noise sources characteristic of Q-meter-based systems, these models substantially reduce fitting-related uncertainties that have historically limited the precision and stability of polarization extraction. The resulting improvements translate directly into enhanced robustness for both real-time (online) monitoring and high-precision offline analysis, which can improve the effective figure of merit by reducing analysis-induced uncertainty.

Across a broad polarization range, the proposed AI-based polarization and area models outperform traditional TE- and lineshape-based methods under realistic noise and baseline conditions, particularly in regimes where conventional fitting becomes unstable or biased. In parallel, the denoising autoencoder provides an effective mechanism for suppressing stochastic noise while preserving physically meaningful spectral structure, enabling downstream extraction methods to operate on cleaner, more interpretable signals. Together, these components form a flexible analysis framework that decouples noise suppression, signal interpretation, and polarization inference in a manner not achievable with traditional pipelines.

At the same time, it is important to emphasize that these methods do not circumvent fundamental instrumental constraints. While neural networks can reduce extraction and fitting uncertainties well below the percent level, the ultimate lower bound on achievable polarization precision remains set by intrinsic limitations of the Q-meter system, thermal-equilibrium calibration accuracy, coil--sample coupling, field homogeneity, and target-specific experimental conditions. In this sense, the principal impact of the AI-based approach is the elimination of avoidable analysis-induced error, thereby allowing the total uncertainty to be dominated by well-understood physical and instrumental effects.

From an operational standpoint, the millisecond-scale inference times achieved by all models represent a significant advantage for online monitoring and adaptive system control. Unlike traditional Dulya-type or area-based fits---which are computationally intensive and often require manual intervention---AI-based inference enables autonomous, high-rate polarization estimation suitable for feedback-driven optimization during data taking. This capability is particularly relevant for experiments operating under rapidly changing conditions or constrained beam-time schedules.

Looking forward, several avenues exist for further refinement. Improving performance in the extreme low-polarization and high-noise regimes will require targeted architectural optimization and expanded training datasets. In particular, neuron-pruning and model-compression strategies offer a promising route to reducing model complexity while simultaneously improving inference speed and generalization. In parallel, continued development of the denoising autoencoder---especially its integration with polarization and area models---will further strengthen the global analysis pipeline.

Finally, the methodology presented here is not limited to a single target material or spin species. Extending the area-based framework to encompass both spin-$\tfrac{1}{2}$ and spin-1 systems within a unified model will enhance adaptability during experimental target changes and facilitate broader deployment across polarized-target programs. Taken together, these results establish deep-learning--assisted CW-NMR polarimetry as a viable, high-performance tool for next-generation polarized scattering experiments.

\section{Acknowledgments}
The authors acknowledge Research Computing at the University of Virginia for providing computational resources and technical support that have contributed to the results reported in this publication.
For additional information, see \url{https://rc.virginia.edu}.
This work was supported by the U.S. Department of Energy under contract DE-FG02-96ER40950.

\bibliographystyle{unsrt}
\bibliography{biblio.bib}

@article{Court1993,
   author = {G R Court and D W Gifford and P Harrison and W G Heyes and M A Houlden},
   doi = {https://doi.org/10.1016/0168-9002(93)91047-Q},
   issn = {0168-9002},
   issue = {3},
   journal = {Nuclear Instruments and Methods in Physics Research Section A: Accelerators, Spectrometers, Detectors and Associated Equipment},
   pages = {433-440},
   title = {A high precision Q-meter for the measurement of proton polarization in polarised targets},
   volume = {324},
   url = {https://www.sciencedirect.com/science/article/pii/016890029391047Q},
   year = {1993},
}

@book{abragam_principles_1986,
	location = {Oxford},
	edition = {Repr},
	title = {The principles of nuclear magnetism},
	isbn = {9780198512363},
	series = {The international series of monographs on physics},
	pagetotal = {599},
	publisher = {Clarendon Pr},
	author = {Abragam, Anatole},
	year = {1986},
}

@article{Keller2017,
   author = {D. Keller},
   doi = {10.1140/epja/i2017-12344-0},
   issn = {1434601X},
   issue = {7},
   journal = {European Physical Journal A},
   title = {Modeling alignment enhancement for solid polarized targets},
   volume = {53},
   year = {2017},
}

@article{Keller2020,
   author = {D. Keller and D. Crabb and D. Day},
   doi = {10.1016/j.nima.2020.164504},
   issn = {01689002},
   journal = {Nuclear Instruments and Methods in Physics Research, Section A: Accelerators, Spectrometers, Detectors and Associated Equipment},
   title = {Enhanced tensor polarization in solid-state targets},
   volume = {981},
   year = {2020},
}

@article{Dulya1997,
   author = {C. Dulya and others},
   doi = {10.1016/S0168-9002(97)00317-3},
   issn = {01689002},
   issue = {2-3},
   journal = {Nuclear Instruments and Methods in Physics Research, Section A: Accelerators, Spectrometers, Detectors and Associated Equipment},
   title = {A line-shape analysis for spin-1 NMR signals},
   volume = {398},
   year = {1997},
}

@article{Abragam1961,
   author = {A. Abragam and L. C. Hebel},
   doi = {10.1119/1.1937646},
   issn = {0002-9505},
   issue = {12},
   journal = {American Journal of Physics},
   title = {The Principles of Nuclear Magnetism},
   volume = {29},
   year = {1961},
}

@article{CLEMENT2023168177,
title = {Manipulation of spin-1 solid-state targets},
journal = {Nuclear Instruments and Methods in Physics Research Section A: Accelerators, Spectrometers, Detectors and Associated Equipment},
volume = {1050},
pages = {168177},
year = {2023},
issn = {0168-9002},
doi = {https://doi.org/10.1016/j.nima.2023.168177},
url = {https://www.sciencedirect.com/science/article/pii/S0168900223001675},
author = {Joseph Clement and Dustin Keller}
}

@article{Hornik1989,
title = {Multilayer feedforward networks are universal approximators},
journal = {Neural Networks},
volume = {2},
number = {5},
pages = {359-366},
year = {1989},
issn = {0893-6080},
doi = {https://doi.org/10.1016/0893-6080(89)90020-8},
url = {https://www.sciencedirect.com/science/article/pii/0893608089900208},
author = {Kurt Hornik and Maxwell Stinchcombe and Halbert White}
}

@inproceedings{CourtNMRSimulation,
  title     = {Modeling Non-Constant Current Effects for a Series Tune NMR Q-Meter Used for Nucleon Polarization Measurements},
  author    = {G.R. Court and M.A. Houlden},
  year      = {1998},
  address   = {University of Virginia, Charlottesville, VA, U.S.A.},
  booktitle = {Proceedings of the Workshop on NMR in Polarized Targets},
  pages = {35-45}
}

@book{hastie01statisticallearning,
  address = {New York, NY, USA},
  author = {Hastie, Trevor and Tibshirani, Robert and Friedman, Jerome},
  publisher = {Springer New York Inc.},
  series = {Springer Series in Statistics},
  title = {The Elements of Statistical Learning},
  year = 2001
}

@inproceedings{Pentilla1998,
  author    = {S. I. Pentill{\"a}},
  title     = {Workshop on NMR in Polarization Targets},
  booktitle = {Proceedings of the Workshop on NMR in Polarized Targets},
  publisher = {University of Virginia},
  address   = {Charlottesville, VA},
  year      = {1998},
  pages     = {15--26}
}

@article{KELLER2013133,
title = {Uncertainty minimization in NMR measurements of dynamic nuclear polarization of a proton target for nuclear physics experiments},
journal = {Nuclear Instruments and Methods in Physics Research Section A: Accelerators, Spectrometers, Detectors and Associated Equipment},
volume = {728},
pages = {133-144},
year = {2013},
issn = {0168-9002},
doi = {https://doi.org/10.1016/j.nima.2013.06.103},
url = {https://www.sciencedirect.com/science/article/pii/S0168900213009625},
author = {D. Keller}
}

@misc{kumar2023imagedataaugmentationapproaches,
      title={Image Data Augmentation Approaches: A Comprehensive Survey and Future directions}, 
      author={Teerath Kumar and Alessandra Mileo and Rob Brennan and Malika Bendechache},
      year={2023},
      eprint={2301.02830},
      archivePrefix={arXiv},
      primaryClass={cs.CV},
      url={https://arxiv.org/abs/2301.02830}, 
}

@inproceedings{inproceedings,
  author = {LeCun, Yann and Kavukcuoglu, Koray and Farabet, Cl{\'{e}}ment},
  booktitle = {Circuits and Systems (ISCAS), Proceedings of 2010 IEEE International Symposium on},
  organization = {IEEE},
  pages = {253--256},
  title = {{Convolutional networks and applications in vision}},
  year = 2010
}

@book{Venkatesan,
title = "Convolutional Neural Networks in Visual Computing: A Concise Guide",
author = "Ragav Venkatesan and Baoxin Li",
note = "Publisher Copyright: {\textcopyright} 2018 by Taylor \& Francis Group, LLC. All rights reserved.",
year = "2017",
month = oct,
day = "23",
doi = "10.4324/9781315154282",
language = "English (US)",
isbn = "9781498770392",
publisher = "CRC Press",
}

@misc{loshchilov2019decoupledweightdecayregularization,
      title={Decoupled Weight Decay Regularization}, 
      author={Ilya Loshchilov and Frank Hutter},
      year={2019},
      eprint={1711.05101},
      archivePrefix={arXiv},
      primaryClass={cs.LG},
      url={https://arxiv.org/abs/1711.05101}, 
}

@misc{he2015deepresiduallearningimage,
      title={Deep Residual Learning for Image Recognition}, 
      author={Kaiming He and Xiangyu Zhang and Shaoqing Ren and Jian Sun},
      year={2015},
      eprint={1512.03385},
      archivePrefix={arXiv},
      primaryClass={cs.CV},
      url={https://arxiv.org/abs/1512.03385}, 
}

@article{shorten2019survey,
  author = {Shorten, Connor and Khoshgoftaar, Taghi M},
  journal = {Journal of Big Data},
  number = 1,
  pages = {1--48},
  publisher = {Springer},
  title = {A survey on image data augmentation for deep learning},
  volume = 6,
  year = 2019
}

@book{vapnik1998statistical,
  title={Statistical Learning Theory},
  author={Vapnik, Vladimir N},
  year={1998},
  publisher={Wiley},
  address={New York},
  isbn={978-0-471-03003-4}
}

@article{zhang2021survey,
  title={Data Augmentation for Deep Learning: A Survey},
  author={Zhang, Qingyun and Yang, Lihua and Zhang, Yidan and Hanne, Thomas},
  journal={Journal of Big Data},
  volume={8},
  number={1},
  pages={1--48},
  year={2021},
  publisher={Springer},
  doi={10.1186/s40537-021-00499-4}
}

@book{goodfellow2016deep,
  author = {Goodfellow, Ian and Bengio, Yoshua and Courville, Aaron},
  note = {Book in preparation for MIT Press},
  publisher = {MIT Press},
  title = {Deep Learning},
  url = {http://www.deeplearningbook.org},
  year = 2016
}

@inproceedings{autoencoder-vincent,
  author = {Vincent, Pascal and Larochelle, Hugo and Bengio, Yoshua and Manzagol, Pierre-Antoine},
  booktitle = {ICML},
  editor = {Cohen, William W. and McCallum, Andrew and Roweis, Sam T.},
  ee = {https://www.wikidata.org/entity/Q57257952},
  isbn = {978-1-60558-205-4},
  pages = {1096-1103},
  publisher = {ACM},
  series = {ACM International Conference Proceeding Series},
  title = {Extracting and composing robust features with denoising autoencoders.},
  url = {http://dblp.uni-trier.de/db/conf/icml/icml2008.html#VincentLBM08},
  volume = 307,
  year = 2008
}

@misc{michelucci2022introductionautoencoders,
      title={An Introduction to Autoencoders}, 
      author={Umberto Michelucci},
      year={2022},
      eprint={2201.03898},
      archivePrefix={arXiv},
      primaryClass={cs.LG},
      url={https://arxiv.org/abs/2201.03898}, 
}

@article{2019JCoPh.378..686R,
title = {Physics-informed neural networks: A deep learning framework for solving forward and inverse problems involving nonlinear partial differential equations},
journal = {Journal of Computational Physics},
volume = {378},
pages = {686-707},
year = {2019},
issn = {0021-9991},
doi = {https://doi.org/10.1016/j.jcp.2018.10.045},
url = {https://www.sciencedirect.com/science/article/pii/S0021999118307125},
author = {M. Raissi and P. Perdikaris and G.E. Karniadakis}
}

@misc{stolte2023dominodomainawarelossregularization,
      title={DOMINO++: Domain-aware Loss Regularization for Deep Learning Generalizability}, 
      author={Skylar E. Stolte and Kyle Volle and Aprinda Indahlastari and Alejandro Albizu and Adam J. Woods and Kevin Brink and Matthew Hale and Ruogu Fang},
      year={2023},
      eprint={2308.10453},
      archivePrefix={arXiv},
      primaryClass={cs.CV},
      url={https://arxiv.org/abs/2308.10453}, 
}

@misc{bronstein2021geometricdeeplearninggrids,
      title={Geometric Deep Learning: Grids, Groups, Graphs, Geodesics, and Gauges}, 
      author={Michael M. Bronstein and Joan Bruna and Taco Cohen and Petar Veličković},
      year={2021},
      eprint={2104.13478},
      archivePrefix={arXiv},
      primaryClass={cs.LG},
      url={https://arxiv.org/abs/2104.13478}, 
}

@misc{szegedy2014goingdeeperconvolutions,
      title={Going Deeper with Convolutions}, 
      author={Christian Szegedy and Wei Liu and Yangqing Jia and Pierre Sermanet and Scott Reed and Dragomir Anguelov and Dumitru Erhan and Vincent Vanhoucke and Andrew Rabinovich},
      year={2014},
      eprint={1409.4842},
      archivePrefix={arXiv},
      primaryClass={cs.CV},
      url={https://arxiv.org/abs/1409.4842}, 
}

@misc{paszke2019pytorchimperativestylehighperformance,
      title={PyTorch: An Imperative Style, High-Performance Deep Learning Library}, 
      author={Adam Paszke and Sam Gross and Francisco Massa and Adam Lerer and James Bradbury and Gregory Chanan and Trevor Killeen and Zeming Lin and Natalia Gimelshein and Luca Antiga and Alban Desmaison and Andreas Köpf and Edward Yang and Zach DeVito and Martin Raison and Alykhan Tejani and Sasank Chilamkurthy and Benoit Steiner and Lu Fang and Junjie Bai and Soumith Chintala},
      year={2019},
      eprint={1912.01703},
      archivePrefix={arXiv},
      primaryClass={cs.LG},
      url={https://arxiv.org/abs/1912.01703}, 
}

@misc{akiba2019optunanextgenerationhyperparameteroptimization,
      title={Optuna: A Next-generation Hyperparameter Optimization Framework}, 
      author={Takuya Akiba and Shotaro Sano and Toshihiko Yanase and Takeru Ohta and Masanori Koyama},
      year={2019},
      eprint={1907.10902},
      archivePrefix={arXiv},
      primaryClass={cs.LG},
      url={https://arxiv.org/abs/1907.10902}, 
}

@article{Crabb1997,
  author       = {D. G. Crabb and W. Meyer},
  title        = {Solid polarized targets for nuclear and particle physics experiments},
  journal      = {Annual Review of Nuclear and Particle Science},
  volume       = {47},
  pages        = {67--109},
  year         = {1997},
  doi          = {10.1146/annurev.nucl.47.1.67}
}

@misc{loshchilov2017sgdrstochasticgradientdescent,
      title={SGDR: Stochastic Gradient Descent with Warm Restarts}, 
      author={Ilya Loshchilov and Frank Hutter},
      year={2017},
      eprint={1608.03983},
      archivePrefix={arXiv},
      primaryClass={cs.LG},
      url={https://arxiv.org/abs/1608.03983}, 
}

@article{LeNail2019, doi = {10.21105/joss.00747}, url = {https://doi.org/10.21105/joss.00747}, year = {2019}, publisher = {The Open Journal}, volume = {4}, number = {33}, pages = {747}, author = {LeNail, Alexander}, title = {NN-SVG: Publication-Ready Neural Network Architecture Schematics}, journal = {Journal of Open Source Software} }

@misc{hu2019squeezeandexcitationnetworks,
      title={Squeeze-and-Excitation Networks}, 
      author={Jie Hu and Li Shen and Samuel Albanie and Gang Sun and Enhua Wu},
      year={2019},
      eprint={1709.01507},
      archivePrefix={arXiv},
      primaryClass={cs.CV},
      url={https://arxiv.org/abs/1709.01507}, 
}

@misc{vaswani2023attentionneed,
      title={Attention Is All You Need}, 
      author={Ashish Vaswani and Noam Shazeer and Niki Parmar and Jakob Uszkoreit and Llion Jones and Aidan N. Gomez and Lukasz Kaiser and Illia Polosukhin},
      year={2023},
      eprint={1706.03762},
      archivePrefix={arXiv},
      primaryClass={cs.CL},
      url={https://arxiv.org/abs/1706.03762}, 
}

@inproceedings{Kielhorn,
    author = {William Frederick Kielhorn},
    title = {A technique for Measurement of Vector and Tensor Polarization in Solid Spin One Polarized Targets} ,
    booktitle = {Proceedings of the 4th International Workshop on Polarized Target Materials and Techniques},
    year = {1984}
}

@inproceedings{MeyerSchilling,
    author = {W. Meyer and E. Schilling},
    title = {Tensor Polarized Deuteron Target for Intermediate Energy Physics},
    booktitle = {Proceedings of the 4th International workshop on Polarized Target Materials and Techniques},
    year = {1984}
}

@article{Adams1999,
  author        = {Adams, D. and Adeva, B. and Arik, E. and others},
  title         = {The polarized double cell target of the {SMC}},
  journal       = {Nuclear Instruments and Methods in Physics Research Section A: Accelerators, Spectrometers, Detectors and Associated Equipment},
  volume        = {437},
  number        = {1},
  pages         = {23--67},
  year          = {1999},
  doi           = {10.1016/S0168-9002(99)00582-3},
  url           = {https://doi.org/10.1016/S0168-9002(99)00582-3},
  note          = {Spin Muon Collaboration (SMC)}
}

@article{Adeva1994,
  author        = {Adeva, B. and Ahmad, S. and Arvidson, A. and others},
  title         = {Measurement of the deuteron polarization in a large target},
  journal       = {Nuclear Instruments and Methods in Physics Research Section A: Accelerators, Spectrometers, Detectors and Associated Equipment},
  volume        = {349},
  number        = {2--3},
  pages         = {334--344},
  year          = {1994},
  doi           = {10.1016/0168-9002(94)91197-5},
  url           = {https://doi.org/10.1016/0168-9002(94)91197-5},
  note          = {Spin Muon Collaboration (SMC)}
}

@article{Dhawan1996,
  author        = {Dhawan, S. K. and Crabb, D. G. and Hayashi, N. and Rijllart, A.},
  title         = {Precision deuteron {NMR} signal measurement with the {NA47} polarized target},
  journal       = {IEEE Transactions on Nuclear Science},
  volume        = {43},
  number        = {3Pt3},
  pages         = {2128--2134},
  year          = {1996},
  month         = jun,
  doi           = {10.1109/23.502306},
  url           = {https://doi.org/10.1109/23.502306}
}

@article{Reicherz2024,
  author        = {Reicherz, Gerhard and Doshita, Norihito and Takanashi, Yuya and Iwata, Takahiro},
  title         = {Target Offline Polarization {COMPASS} 2022},
  journal       = {PoS},
  volume        = {SPIN2023},
  pages         = {202},
  year          = {2024},
  doi           = {10.22323/1.456.0202},
  url           = {https://pos.sissa.it/456/202/},
  note          = {On behalf of the COMPASS Polarized Target Group}
}

@article{Adeva1998AmmoniaEST,
  author       = {Adeva, B. and others},
  collaboration = {Spin Muon Collaboration},
  title        = {Measurement of proton and nitrogen polarization in ammonia and a test of equal spin temperature},
  journal      = {Nuclear Instruments and Methods in Physics Research Section A: Accelerators, Spectrometers, Detectors and Associated Equipment},
  volume       = {419},
  number       = {1},
  pages        = {60--82},
  year         = {1998},
  doi          = {10.1016/S0168-9002(98)00916-4}
}

@article{Kramer1995HighPrecision,
  author       = {Kr{\"a}mer, D. and others},
  title        = {High precision measurement of the polarization in a large target},
  journal      = {Nuclear Instruments and Methods in Physics Research Section A: Accelerators, Spectrometers, Detectors and Associated Equipment},
  volume       = {356},
  pages        = {79--82},
  year         = {1995},
  doi          = {10.1016/0168-9002(94)01448-5}
}

@article{Adams1997SMCProtonG1,
  author       = {Adams, D. and others},
  collaboration = {Spin Muon Collaboration},
  title        = {Spin structure of the proton from polarized inclusive deep-inelastic muon-proton scattering},
  journal      = {Physical Review D},
  volume       = {56},
  pages        = {5330--5358},
  year         = {1997},
  doi          = {10.1103/PhysRevD.56.5330},
  eprint       = {hep-ex/9702005},
  archivePrefix = {arXiv}
}

\end{document}